\documentclass[aps,
longbibliography,notitlepage,reprint,superscriptaddress]{revtex4-2}
\usepackage[colorlinks,linkcolor=blue,citecolor=blue,urlcolor=red]{hyperref}
\usepackage{soul}
\usepackage{changes}
\usepackage{float}
\usepackage{CJK}
\usepackage{graphicx}%,float}
\usepackage{amsmath,amssymb,amsfonts,physics}
\usepackage{siunitx}
\usepackage{soul}
\usepackage{chngcntr}
\usepackage{chemformula}
\usepackage{MnSymbol}
\usepackage{dsfont} % proper mathbb format
\usepackage{appendix}

\makeatletter

\def\@author@join@#1#2#3{%
  \def\@author{{#1}{#2\revtex@separate@note{#3}}}%
}

\def\revtex@separate@note#1{%
  \comma@space\frontmatter@footnote{#1}%
}

\def\doauthor#1#2#3{%
  \ignorespaces#1\unskip\@listcomma
  \begingroup
    #3%
    \@if@empty{#2}{\endgroup{}{}}{\endgroup{}{}#2}%
  \space\@listand
}

\makeatother

\begin{document}

\raggedbottom
\interfootnotelinepenalty=10000	
\newcommand{\bd}[1]{\mathbf{#1}}
\newcommand{\SiN}{$\mathrm{Si_3N_4}$}
\newcommand{\HU}{John A. Paulson School of Engineering and Applied Sciences, Harvard University, Cambridge, MA 02138, USA}

\author{Guanhao Huang}
\thanks{These authors contributed equally to this work.}
\email{guanhao\_huang@seas.harvard.edu}
\affiliation{\HU}
\author{Amberly Xie}
\thanks{These authors contributed equally to this work.}
\affiliation{\HU}
\author{Chang Jin}
\affiliation{\HU}
\author{Evelyn Hu}
\affiliation{\HU}
\author{Marko Lončar}
\email{loncar@seas.harvard.edu}
\affiliation{\HU}

\title{Enhanced SiV--phonon coupling in dissipation-diluted diamond nanomechanics}

\begin{abstract}
Solid-state color centers embedded into mechanical resonators, where orbital states mediate interactions between spins and long-lived phonons, are a promising platform for hybrid quantum systems. The negatively charged silicon vacancy center in diamond (SiV\textsuperscript{-}) is particularly favorable for strain-coupling to nanomechanical resonators because of its large spin-strain susceptibility and compatibility with device integration. Tensioned nanomechanical resonators offer an attractive route due to their ultra-long mechanical coherence times enabled by diluting defect-induced dissipation, though it is unclear if the same dilution effect undermines strain-mediated coupling to material defect centers. Here, we experimentally demonstrate a tensioned diamond nanobeam platform that enables enhanced strain coupling to SiV\textsuperscript{-} orbital states. 
Through monitoring the optical response of the SiVs to the static and dynamical strain of the mechanical resonator, we observe that dynamical strain localizes near the clamps of these resonators under static strain. This allows us to enhance SiV-mechanical coupling and the mechanical coherence simultaneously, and provide a SiV-strain sensitivity with greatly improved accuracy over previous literature values.
Thus, our work opens a pathway towards seconds-long coherence hybrid quantum systems utilizing point defects in diamond nanomechanics for quantum information processing.
\end{abstract}

\maketitle

% 1st paragraph: hybrid quantum systems for QIP/QC
Achieving scalable and high-fidelity quantum information processing systems requires combining strong quantum interactions with long coherence times, motivating the development of hybrid architectures that leverage the complementary strengths of different platforms~\cite{kurizki2015quantum, xiang2013hybrid}. Combining disparate systems, though, can introduce additional sources of decoherence that undermine these advantages. 
For example, intrinsic material defects and interfacial losses associated with heterogeneous material integration often degrade the quantum coherence of both systems~\cite{yang2023piezoelectric, gruenke2025surface}. Thus, developing hybrid architectures that maintain strong interactions without compromising coherence times remains a central challenge in advancing hybrid quantum technologies. 

% 2nd paragraph: spin mechanics
Spin-mechanical systems based on solid-state color centers, in which electronic spins couple to mechanical excitations~\cite{Rabl2010}, offer a promising route towards building hybrid quantum systems. For instance, acoustic systems can exhibit exceptionally long coherence times~\cite{Youssefi2023}, which can be further improved with techniques such as dissipation dilution~\cite{engelsen2024ultrahigh, fedorov2019generalized}. To this end, diamond offers exceptionally low intrinsic mechanical dissipation~\cite{huang2025ultracoherent}, as a result of its high crystalline quality and low defect density. Furthermore, among solid-state point defects, the negatively charged silicon vacancy center in diamond (SiV$^-$, referred to as SiV) is particularly well-suited for interfacing to mechanics because of its exceptionally large spin-strain response~\cite{meesala2018strain, sohn2018controlling}, enabling strong coupling to mechanical excitations. Together with the SiV's inversion symmetry that leads to reduced sensitivity to surface electric-field fluctuations, this approach allows integration of spin and mechanical degrees of freedom within a monolithic platform, avoiding many of the challenges associated with heterogeneous material integration~\cite{yang2026mechanical, chu2017quantum, Fung24, Arabmoheghi2026}.

\begin{figure*}
    \centering
    \includegraphics[width=1\linewidth]{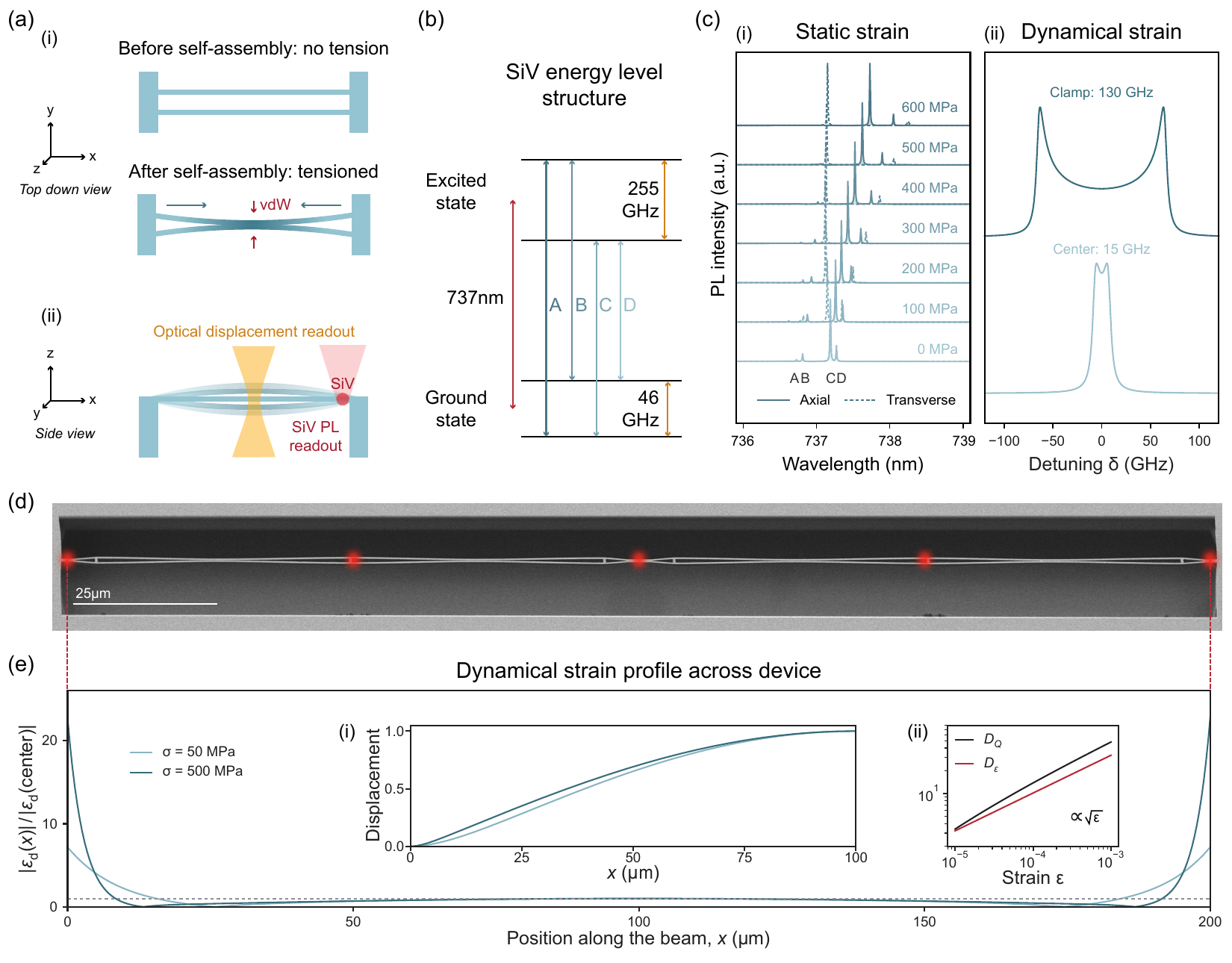}
    \caption{Enhanced SiV-mechanical coupling in a vdW-tensioned diamond nanomechanical resonator. \textbf{(a)} (i) Two diamond nanobeams are brought together with a vdW self-assembly process~\cite{huang2025ultracoherent} utilizing water condensation. (ii) The displacement of the resonator devices is read out using optical interferometry. SiVs at the device clamp experience enhanced defect-phonon coupling as a result of localized dynamical strain. \textbf{(b)} Electronic level structure of the SiV defect center at zero strain showing ground and excited manifolds with spin-orbit eigenstates. The four optical transitions A, B, C, and D are indicated. \textbf{(c)} SiV photoluminescence (PL) spectra as a function of uniaxial [110] static (i) and dynamical strain (ii). Static strain changes the electronic energy level manifold, leading to the shifting of the zero phonon lines (ZPL) of the transverse (90\textdegree to the beam axis) and axial (35\textdegree) SiVs differently. ZPL frequency modulation manifests under dynamical strain, whose amplitudes depend on SiV locations along the beam. \textbf{(d)} SEM of a diamond nanomechanical resonator with SiV implantation spots highlighted in red. \textbf{(e)}  Normalized dynamical strain $\epsilon_d$ profile across the length of the device at two different static stress $\sigma$. As the $\sigma$ increases, the dynamical strain is localized more at the clamp region. Insets: (i) Displacement as a function of position across the device. (ii) The dilution factors $D_Q$ for mechanical quality factors and the enhancement factors $D_{\epsilon}$ for dynamical strain as a function of static strain $\epsilon$. }
    \label{Fig1}
\end{figure*}

% 3rd paragraph: how did we realize this
Here, we realize this approach by integrating SiVs into self-tensioned, dissipation-diluted mechanical resonators~\cite{huang2025ultracoherent} (Fig.~\ref{Fig1}a). We find that high-energy ion implantation needed for SiV generation does not measurably degrade the intrinsic material quality of the resonators, demonstrating the compatibility of defect integration with high-Q mechanical resonators and overcoming many of the challenges associated with heterogeneous hybrid system integration. Additionally, we find that the suppression of mechanical decoherence afforded by dilution of defect-induced dissipation does not come at the expense of strain-mediated SiV-mechanical coupling. Rather, dynamical strain localization at clamped regions enhances SiV coupling to strain~\cite{fedorov2019generalized} (Fig.~\ref{Fig1}e). These findings establish a route to simultaneously engineering long mechanical coherence times while enhancing strong defect-phonon interactions within a single monolithic device.

% 4th paragraph: need to understand in detail how strain couples to SiV and transition to DC strain
A quantitative understanding of such defect-mechanical interactions requires accurate knowledge of the SiV strain susceptibilities that relate local lattice deformations to changes in the SiV energy-level manifold. Prior measurements of these susceptibilities~\cite{meesala2018strain, sohn2018controlling} have relied primarily on electrostatic actuation of cantilever-based geometries with individual SiVs. In such devices, the strain experienced by the SiV depends heavily on its implantation depth and precise knowledge of the electromechanical properties of the heterogeneous device, which can be difficult to accurately model given the complexities of metal deposition, thermal cycling, and surface charge dynamics. Together, these factors limit the reliability with which SiV-mechanical coupling can be predicted in nanoscale devices.

To obtain more accurate measurements of the SiV strain sensitivity, here, we experimentally characterize its optical response to static strain using the self-tensioned nanomechanical resonators (Fig.~\ref{Fig1}a) containing low-density SiV ensembles, thereby greatly suppressing systematic uncertainties. One critical improvement we demonstrate is in using the resonator frequency as a direct measure of the tensile stress of the nanobeam, which relies on only intrinsic material properties of diamond that are well understood. As a result, we obtain revised values for the SiV strain susceptibilities, which are substantially lower than those previously reported~\cite{meesala2018strain, sohn2018controlling}. Thus, these findings further highlight the importance of the demonstrated external enhancement of the defect-strain coupling and enable more accurate predictions regarding the performance reach of monolithic hybrid systems that combine strong defect-phonon interactions with long-lived acoustic memories. 

\section*{Static strain tuning of SiV using vdW self-assembled resonators}

% figure uses PL/20260226/20260226_L60_w02_D3_spot3_GreenMAX_40sec_expo device
\begin{figure}[h!]
    \centering
    \includegraphics[width=1\linewidth]{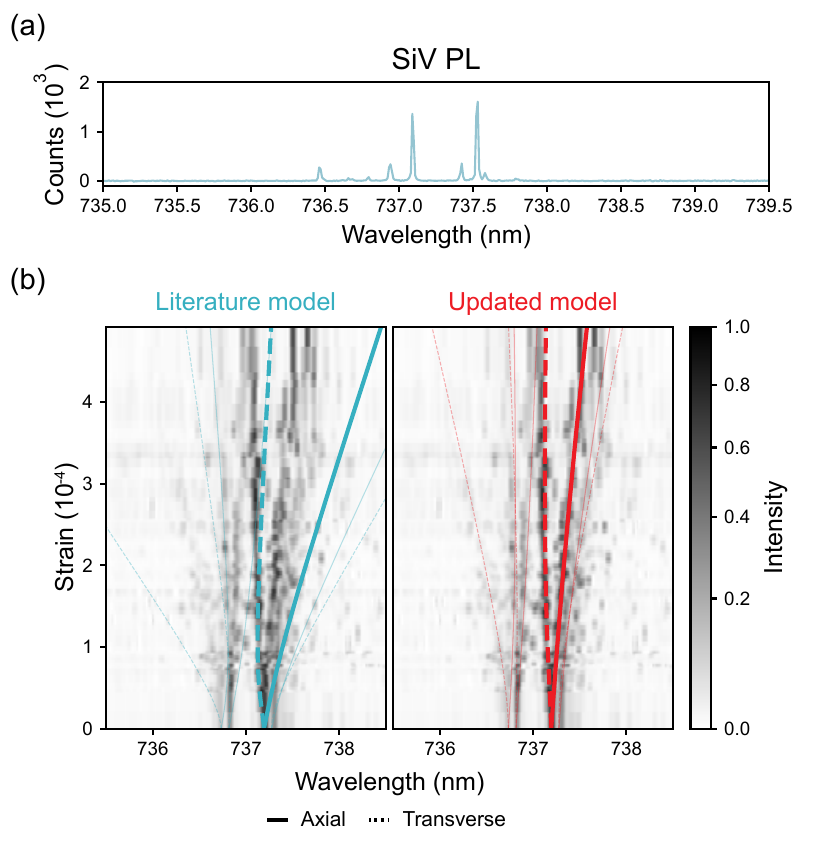}
    \caption{Accurate characterization of SiV strain susceptibility using self-tensioned diamond resonators.  \textbf{(a)} Representative PL spectrum at \SI{4}{K} of SiVs under 0.24 GPa of stress ($2.29 \times 10^{-4}$ strain). \textbf{(b)} Measured SiV PL spectra as a function of static strain. Overlaid theory lines use susceptibility parameters from Ref.~\cite{meesala2018strain, sohn2018controlling} (left) and our updated parameters (right). C-line is bolded for easier visualization. See SI section D on details of strain corrections due to local variations in thickness.}
    \label{Fig2}
\end{figure}

% 1st paragraph topic: why is static strain control important for SiV?
As noted earlier, the SiV is one of the most strain-sensitive color centers, making static strain a powerful tool for controlling its optical and spin properties. In particular, uniaxial strain along the [110] axis has been shown to split the ground and excited electronic energy levels, producing distinct spectral responses for axial and transverse oriented SiVs (Fig.~\ref{Fig1}c)~\cite{meesala2018strain, assumpcao2023deterministic}. This tunability has been used to bring spectrally distinct emitters, whose optical characteristics depend on the local lattice environment and thus vary from emitter to emitter, into resonance, facilitating entanglement between disparate SiVs~\cite{machielse2019quantum}. Strain engineering can also increase the ground state splitting (Fig.~\ref{Fig1}b), which has been demonstrated to extend the SiV spin-coherence time by suppressing phonon-mediated relaxation pathways~\cite{sohn2018controlling}, and offers a potential route in extending SiV operation to higher temperatures. 

% 2nd paragraph: overview/background of what we did and why it's better again than what was previously done
To experimentally quantify the SiV strain response, we incorporate them into vdW-tensioned nanomechanical resonators~\cite{huang2025ultracoherent}, which exhibit geometry-dependent tension (SI section A for device fabrication). A representative SEM of the self-tensioned strings is shown in Fig.~\ref{Fig1}d, with length $L=\SI{200}{\micro\meter}$ and thickness $h=\SI{217}{nm}$. Owing to their large aspect ratios (\SI{e3}{}) and structural simplicity, the resonators can be well approximated to a 1D tensioned string, such that the measured mechanical resonance frequency $f=(1/2L)\sqrt{\sigma/\rho}$ provides a direct measure of their tensile stress $\sigma$, given the length $L$ of the string and the mass density $\rho$ of diamond. This approach avoids uncertainties associated with previous measurements of the SiV strain sensitivity, which relied solely on simulation of complex material stacks sensitive to the aforementioned physical factors~\cite{meesala2018strain} and hindered accurate knowledge of the device deformation. The doubly-clamped geometry of our devices also produces a strain profile that is uniform across different SiV implantation depths, eliminating a key source of systematic uncertainty associated with implantation straggle in cantilever-based devices whose axial strain across the depth shows a strong gradient. Lastly, measurements of low-density ensembles (3-5 SiVs) capture the average strain response across multiple SiVs rather than relying on individual defect responses susceptible to local lattice disorder. The combination of these factors enables a more accurate determination of the SiV strain response than has previously been accessible. 

% 3rd paragraph: our findings and why they are important
SiVs were implanted at 5 locations along the beam, enabling the SiV strain response to be probed along the length of the resonator. Mapping the calibrated strain at the SiV locations with their photoluminescence (PL) spectra reveals a clear stress-dependent relationship (Fig.~\ref{Fig2}b) that deviates significantly from predictions based on previously reported strain susceptibility parameters~\cite{meesala2018strain}. Fitting the spectral map by multiplying a global factor to the literature susceptibility, in an attempt to account for an unknown systematic error in the prior work, yields substantially improved agreement with the data, indicating an orbital strain response approximately threefold lower (a factor of 0.356) than previously reported. As an independent validation, we additionally extract a similar correction factor from the temperature dependence of the ground state SiV T1 coherence time from Ref.~\cite{kuruma2025controlling} (SI section E). The analysis yields susceptibility parameters approximately 0.37 times smaller than those compared to~\cite{meesala2018strain}, in close agreement with the measurement results presented here. These results establish a substantially weaker SiV strain response than previously assumed, motivating a reassessment of the achievable cooperativity in SiV-based spin-mechanical devices and the development of new techniques to overcome this. 

% took this out: However, we do note that these estimates have other types of systematic uncertainties compared to the prior cantilever work~\cite{meesala2018strain}, such as the complex geometrical corrections due to vdW self-assembly structures in the static strain analysis, as well as the surface-mode contributions in the lifetime analysis. (can maybe put back in if people ask)

\section*{Dynamical strain coupling between SiV and flexural motions}

\begin{figure*}
    \centering
    \includegraphics[width=1\linewidth]{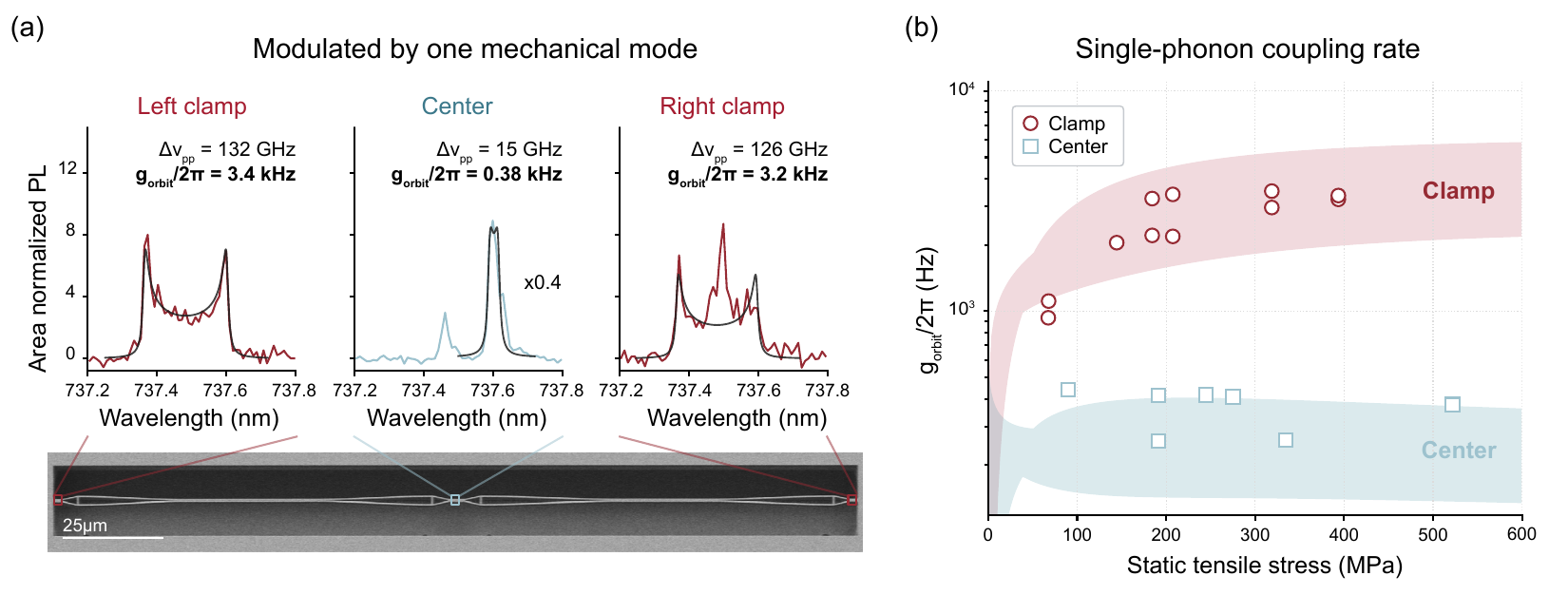}
    \caption{Enhanced SiV orbital-resonator coupling due to dynamical strain localization at the resonator clamp. \textbf{(a)} Time-averaged PL spectra of SiV defect centers implanted at the clamp and the center of a mechanical resonator where the fundamental out-of-plane mode at \SI{810}{kHz} ($\sim \SI{500}{MPa}$ tensioned) is excited. The single-phonon coupling rates $g_{\text{orbit}}$ with the SiV orbitals are characterized using the measured spectral peak splittings $\Delta \nu_{\text{pp}}$ and the resonator displacement amplitudes. Differences between the left and right spectra are attributed to nonlinear dynamics in the driven resonator, which lead to temporal envelope modulation of the dynamical strain. \textbf{(b)} Orbital coupling rates $g_{\text{orbit}}$ as a function of static tensile stress. Static stress enhances the coupling rates at the clamp tenfold compared to those at the center. The shaded region covers possible SiV emission lines with different orientations in the measurement window. }
    \label{Fig3}
\end{figure*}

% can still achieve good mechanical coherence
The reduced SiV strain response identified here highlights the need for strategies that enhance defect–phonon coupling without compromising mechanical coherence. Monolithic integration of solid-state color centers with quantum acoustic devices~\cite{Flühmann2019} offers a promising approach by avoiding coherence degradation pathways that commonly occur for heterogeneous systems~\cite{chu2017quantum, doi:10.1126/science.adr2464, doi:10.1126/science.aeh7535}. In our SiV-integrated diamond resonators, we measure an intrinsic material quality factor of $Q_\mathrm{int}\approx\SI{e6}{}$ at 4K, with no measurable degradation after high-energy ion implantation typically used for SiV formation. Furthermore, the theoretical impact of SiV-induced relaxation damping is also negligible (SI section F). 

% spins coupling to mechanics
While there have been several previous studies of spin-mechanical systems utilizing different coupling mechanisms~\cite{Kolkowitz12,Fung24}, many of these approaches have uncovered excess dissipation~\cite{Arabmoheghi2026}. Our approach instead utilizes vdW-induced tension to simultaneously enhance mechanical coherence through dissipation dilution and strengthen SiV–mechanical coupling through dynamical strain localization, leading to around a thousand-fold improvement in the SiV ground-state cooperativity, reaching $C_\mathrm{GS}=4g_\mathrm{GS}^2/\gamma_\mathrm{GS}\Gamma_m\sim 100$. Dissipation dilution enhances the mechanical quality factor by introducing a lossless restoring potential through static tension to the flexural modes of the device~\cite{engelsen2024ultrahigh}. Consistent with this mechanism, our most highly strained devices exhibit dilution factors up to $D_Q=Q_m/Q_\mathrm{int}\approx30$, reaching mechanical quality factors up to $Q_m=\SI{26e6}{}$ (SI section F). One interesting side effect of this is dynamical strain localization, $\epsilon_\mathrm{d}(x_0)$, at the clamp boundary. The dynamical strain exhibits profile $e^{-x_0/\ell}$, where $x_0$ is the distance to the clamping point and $\ell = h/\sqrt{12\epsilon}$ is the characteristic boundary length~\cite{fedorov2019generalized}. Increasing the static strain, $\epsilon$, therefore amplifies the dynamical strain experienced by the SiV near the clamp, leading to enhanced SiV-mechanical coupling by a factor of $D_\epsilon=L/\pi\ell$, similar to the dissipation dilution factor $D_Q=L/2\ell$. 

We demonstrate this effect by comparing the orbital-phonon coupling rates, $g_\mathrm{orbit}$, of SiVs at the clamp and beam center with varying static strain (Fig.~\ref{Fig3}b). $g_\mathrm{orbit}$ is characterized by measuring the acoustic modulation of the SiV PL spectrum (Fig.~\ref{Fig3}a), and the resonator dynamical displacement amplitude $q$ using optical interferometry (SI section H). Fundamental flexural modes were externally excited to large amplitudes, $q\sim \SI{500}{nm}$, using a piezo mounted on the back of the sample holder. SiVs at the clamp show ten-fold improvement of $g_\mathrm{orbit}$ over those at the center, and only the SiVs at the clamps show coupling enhancement over static strain. The close agreement between the theoretically predicted values and our measurements further validates the correction of SiV strain sensitivity and demonstrates its importance in predicting SiV-mechanical coupling behavior.

\begin{figure*}
    \centering
    \includegraphics[width=1\linewidth]{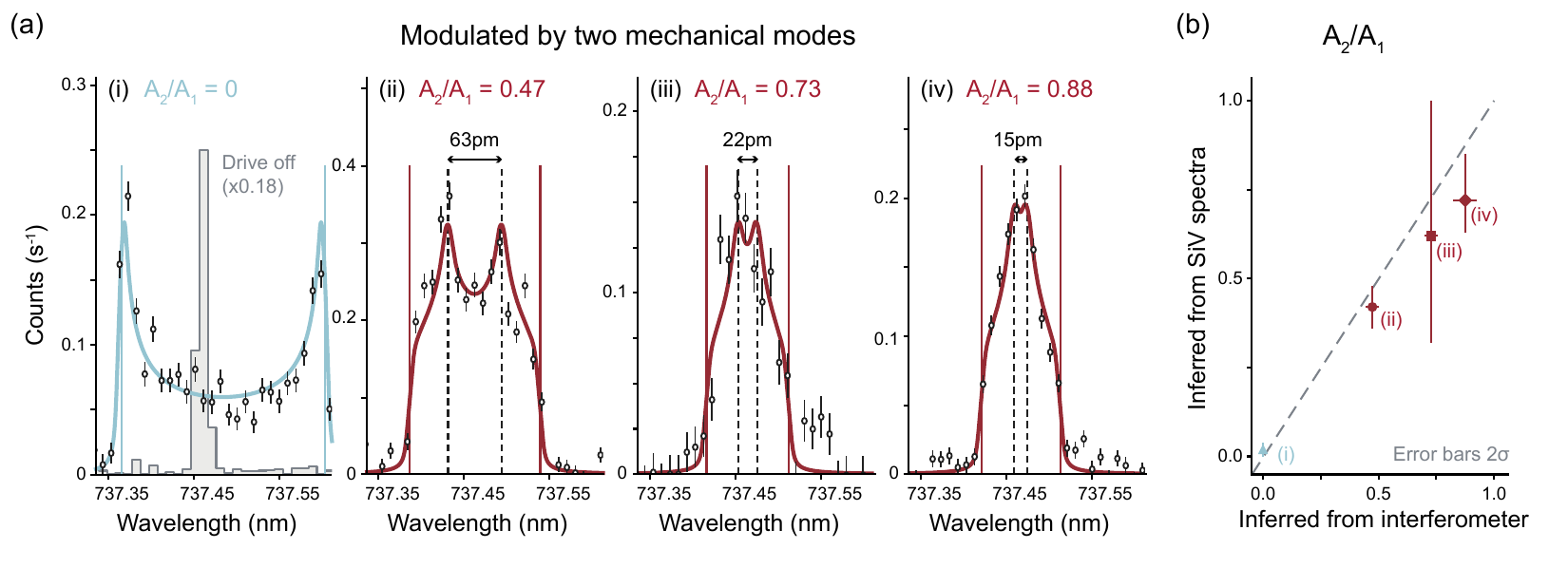}
    \caption{Connection between SiV PL spectra and the multi-frequency phononic excitations. \textbf{(a)} Time-averaged PL spectra of a SiV defect center implanted at the left clamp when both fundamental and second-order out-of-plane modes are externally excited, with PL modulation amplitudes $A_1$ and $A_2$ from each mode. This allows us to study PL spectra under nonlinear acoustic dynamics by mimicking a temporal modulation of the dynamical strain amplitude from $|A_1 + A_2|$ to $|A_1 - A_2|$. The solid vertical lines mark the maximum modulation amplitude $|A_1 + A_2|$, whereas the dashed vertical lines mark the peak splitting set by $|A_1 - A_2|$. The $A_2/A_1$ value at the top of the plot is estimated using the interferometer signal taken 3-\SI{5}{\micro\meter} away from the clamp. \textbf{(b)} Comparison of the modulation amplitude ratios inferred by the interferometer signal (horizontal axis) and fitted from the SiV PL spectra (vertical axis), demonstrating the unique connection between SiV PL spectra and the temporal modulation of the dynamical strain amplitude.}
    \label{Fig4}
\end{figure*}

Additionally, we observe that the SiV PL spectrum sometimes deviates from the characteristic double-peak envelope of single-tone modulation $p(\delta)=1/\pi\sqrt{A^2-\delta^2}$, where $A$ is the modulation amplitude and $\delta$ is the SiV line shift. This suggests that SiV is not only sensitive to harmonic lattice strain modulation, but also to temporal amplitude modulation of vibrations that typically arise from nonlinear acoustic dynamics~\cite{Ochs22}. We find that large displacement amplitudes lead to temporal modulation in the form of acoustic frequency combs, mediated by nonlinear interaction with higher-order flexural modes~\cite{Ganesan17,Qi20} (SI section G). This behavior explains some of the non-ideal modulation shape we observed in Fig.~\ref{Fig3}a, whose peak locations indicate the maximum and minimum amplitudes of the temporal amplitude modulation. 

To study this behavior systematically, we simultaneously excite two flexural modes of the mechanical resonator with dynamical strain amplitudes $A_1$ and $A_2$, mimicking the amplitude temporal modulation from $A_1+A_2$ to $|A_1-A_2|$. Monitoring the SiV PL spectrum allows us to probe the dynamical strain profile locally at the clamp, where the dynamical displacement is essentially zero and inaccessible using direct optical interferometry. We observe similar modifications of SiV PL spectra (Fig.~\ref{Fig4}a), whose envelopes feature the van Hove singularities of a 2D band structure~\cite{Van53} and yield consistent modulation ratios $A_2/A_1$ inferred from optical interferometry signals measured 3-\SI{5}{\micro\meter} away from the clamping point (Fig.~\ref{Fig4}b, SI section G). 

Beyond hybrid quantum systems, these measurements demonstrate that SiVs can uniquely provide a direct strain probe at displacement nodes of acoustic devices, complementary to optical interferometry. SiVs can therefore directly map out dynamical strain-induced losses that arise in engineered mechanical resonators and study acoustic systems with nonlinear dynamics.

\section*{Discussion and outlook}

% first paragraph: one sentence (brief!) summary about what we did and how it fits into the bigger picture
The updated SiV-strain susceptibilities reported in this work indicate that all spin-phonon coupling rates predicted from earlier work~\cite{meesala2018strain} are overestimated by almost a factor of 3, and cooperativity $C$ by nearly an order of magnitude. Furthermore, we show that static strain not only enables ten-fold enhancement in coupling rates, but also dilutes mechanical losses, leading to a thousand-fold improvement in cooperativity.

% These values are based on the static and dynamical SiV strain response measured here, which was further confirmed by ground-state lifetime analysis from~\cite{kuruma2025controlling}. (also took this out, is redundant i feel)

% Compared to GHz implementations, the MHz acoustics studied here further allows parametric coupling gated by external controls~\cite{Kolkowitz12}, enabling programmable memory or bus functionalities~\cite{Bennett13}, where our devices already integrate SiVs at both clamps of the mechanical resonator. (i'm commenting this out because we don't talk about difference in mechanical frequency elsewhere, but can put this back in if reviewers ask, and this part of the discussion is still talking about our work)

While we primarily characterized coupling of the SiV orbital states to the mechanical motion of the resonator, realizing a coherent spin-phonon interface will require coupling to defect spin states. Even though the orbital cooperativity $C_\mathrm{GS}\sim 100$ is already beyond unity, the orbital quantum cooperativity $C_{\mathrm{GS},q}=C_\mathrm{GS}/\overline{n}_{\mathrm{th}}$ is only $10^{-3}$ at \SI{4}{K} in this work, where $\overline{n}_{\mathrm{th}}$ is the thermal occupation of the MHz-oscillator. However, we believe that spin-phonon cooperativities of $C_{\mathrm{spin},q}\sim10^5$ can be achieved with similar device geometries. We expect four orders of magnitude improvement by replacing orbital states with SiV spin states under dynamical decoupling~\cite{Sukachev17} at mK temperatures. Although static strain along the [110] axis suppresses spin–orbit coupling and thus spin–strain sensitivity, alternative clamp geometries can spatially separate static and dynamical strain, preserving strong spin–strain sensitivity while enhancing spin–phonon coupling to $g_\mathrm{spin}/2\pi\sim\SI{1}{kHz}$ (see SI section I).
Another five orders of magnitude improvement is estimated by operating the system at \SI{10}{mK}, where acoustic quality factors beyond $10^{10}$ have been demonstrated with thermal decoherence times beyond \SI{10}{\second}~\cite{huang2025ultracoherent}. Though we did not observe measurable excess loss at 4K due to the SiV implantation, the previously demonstrated material quality factor beyond $10^9$ at \SI{10}{mK} of pristine diamond remains to be validated. 

Therefore, our system is well-suited for encoding error-corrected quantum registers using long-lived quantum acoustic systems coupled to solid-state spins, similar to how trapped ions use motional modes~\cite{Flühmann2019} and superconducting circuits utilize bulk microwave cavities~\cite{Brock2025} as long-lived bosonic registers. The monolithic nature of our platform also avoids the compromises typically occurring for heterogeneous integration, frequently observed for superconducting qubit systems~\cite{chu2017quantum, han2020cavity, yang2026mechanical,yang2023piezoelectric, gruenke2025surface} and spin-mechanical systems~\cite{Kolkowitz12,Fung24,Arabmoheghi2026}. 
Future integration of optical interfaces at locations of high dynamical strain will enable high-fidelity quantum control of SiV spins, which in turn allows for the generation of long-lived quantum acoustic states. Beyond providing a foundation for hybrid quantum systems, these macroscopic quantum states could even explore coherence regimes where self-gravitation from massive superposition states could lead to the breakdown of the linear superposition principle of quantum mechanics~\cite{Penrose2014}. 

\section*{Acknowledgments}
We thank Sophie Weiyi Ding, Srujan Meesala, Slava Dobrovitski, Renee George, Katie Barajas, and Aaron Day for helpful discussions. This work was supported by AFOSR (Grant No. FA9550-23-1-0333), AWS (Grant No. A50791, Grant No. A60290), and G.H. gratefully acknowledges financial support from the Swiss National Science Foundation (Postdoc.Mobility, grant number 222257), and Harvard’s Aramont Fellowship for Emerging Science Research. This work was performed in part at the Harvard University Center for Nanoscale Systems (CNS); a member of the National Nanotechnology Coordinated Infrastructure Network (NNCI), which is supported by the National Science Foundation under NSF award no. ECCS-2025158.

\section*{Contributions}
G.H. conceived the project, designed the devices and the experiment. A.X., G.H. fabricated the devices. A.X., C.J., G.H. performed the measurements. G.H., A.X., C.J. analyzed the data. A.X., G.H., C.J. wrote the manuscript with input from all authors. M.L., G.H. supervised the project. All authors contributed to the discussion and writing of the manuscript. 
% \section{Methods}

\appendixpage
\appendix
\setcounter{tocdepth}{2}
\counterwithin{figure}{section}
% \section{Appendix}

\section{Device fabrication}

\begin{figure}[h!]
    \centering
    \includegraphics[width=1\linewidth]{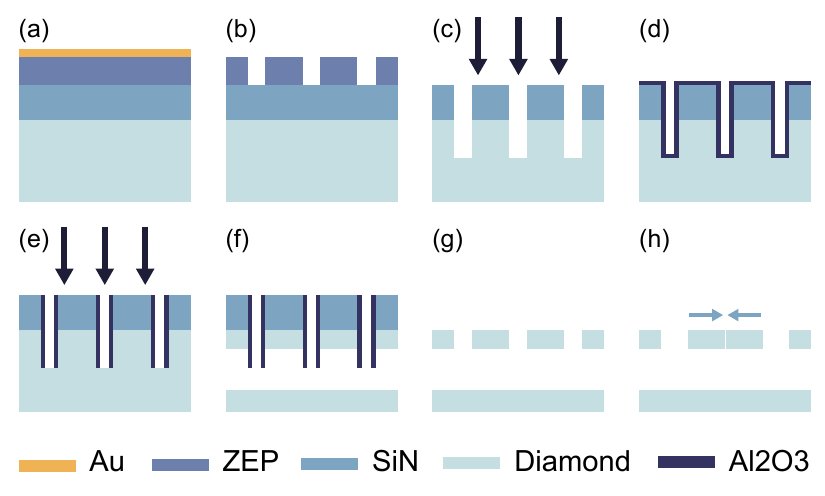}
    \caption{Fabrication process flow for vdW-tensioned diamond resonators. \textbf{(a)} Preparation of the substrate for electron-beam lithography using positive resist. \textbf{(b)} Resist pattern developed. \textbf{(c)} Resist pattern is transferred to the SiN hardmask, then etched into the bulk diamond crystal. \textbf{(d)} Atomic layer deposition (ALD) is used for sidewall protection. \textbf{(e)} Preparation for the quasi-isotropic etching by etching away the top surface of the ALD protection layer. \textbf{(f)}  Quasi-isotropic oxygen plasma etching to undercut the structures. \textbf{(g)} Wet-acid clearning leads to the suspension of high-aspect-ratio slab-like nanomechanical string devices. \textbf{(h)} vdW self-assembly using water vapor leads to tension in diamond strings. }
    \label{FigA1}
\end{figure}

SiV-implanted diamond nanomechanical resonators were fabricated utilizing a quasi-isotropic etching process. Details can be found in various references, including~\cite{kuruma2025controlling,kuruma2021coupling, huang2025ultracoherent} (Fig.~\ref{FigA1}). 
For SiV implantation, electronic-grade diamond samples (Element Six) were masked with MMA/PMMA and written with electron-beam lithography (Elionix F125), preparing for masked spot-implantation. Samples were implanted with Si(29)$^+$ at \SI{90}{keV} and subsequently annealed (\SI{1.5}{\hour} at \SI{1250}{\celsius}, \SI{2e-8}{Pa}), with triacid cleaning performed before and after for removing the graphitization layer. The presence of SiV PL signals were verified before device fabrication in a 4~K cryostat. Samples were further cleaned with 49\% HF, piranha, and solvent. 

Subsequent device fabrication process flow is illustrated in Fig.~\ref{FigA1}. A 200~nm SiN was deposited with PECVD for use as a hard mask to etch diamond, followed by 340~nm ZEP520a, 5~nm Au, and electron beam lithography for device patterning. The SiN hard mask was dry etched using C\textsubscript{4}F\textsubscript{8}, SF\textsubscript{6}, and H\textsubscript{2} plasma, followed by ZEP removal using an overnight \SI{80}{\celsius} PG Remover soak. The SiN pattern was then transferred into the diamond using an oxygen plasma etch targeting a depth of \SI{800}{nm}. To protect the sidewall of the device during the quasi-isotropic undercut etch, an 18~nm Al\textsubscript{2}O\textsubscript{3} ALD layer was deposited, followed by a top-down Ar/Cl\textsubscript{2}  etch to expose the top surface of the diamond trenches. After, the devices were quasi-isotropically undercut with a high-density and high-temperature oxygen plasma until the device thickness was measured around \SI{217}{nm} under SEM. The SiN hard mask and the alumina sidewall layer were removed with acid cleaning. The device was then finished with a 3~nm atomic layer etching to remove plasma-induced surface damage during prior fabrication steps to further improve the mechanical performance of the devices~\cite{huang2025ultracoherent}. After another wet acid cleaning to remove the sputtered Si and Al atoms on the surface, the devices were then tensioned using a van der Waals self-assembly process by passing the devices through water vapor.

\section{SiV optical characterization}
% motivation, methods, what you will see, and the physics behind

\begin{figure}[t!]
    \centering
    \includegraphics[width=1\linewidth]{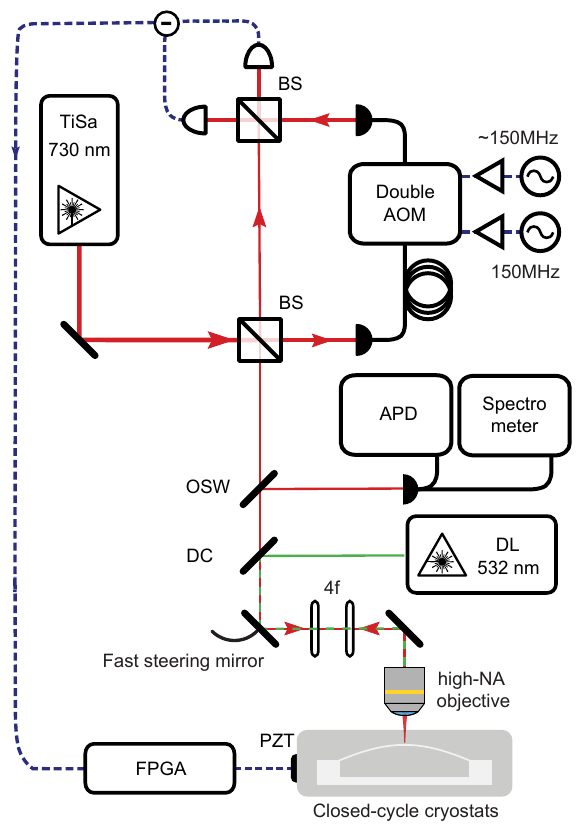}
    \caption{Simplified scheme of the measurement setup combining optical interferometry for mechanical displacement measurements and SiV PL measurements. DL: Diode laser, BS: beam splitter, AOM: acousto-optic modulator, PZT: piezoelectric actuator, FPGA: field-programmable gate array, OSW: optical switch, DC: dichroic mirror, APD: avalanche photodiode.}
    \label{Fig_SI_setup}
\end{figure}

For measuring both SiV PL and mechanical displacement signals, we use the setup shown in Fig.~\ref{Fig_SI_setup}. Devices were glued onto a sample holder with silver paste, and then mounted into an Attodry cryostat (4K). To analyze the SiV emission line spectra as a function of static strain, we excite the SiVs using a 4f confocal setup with a 532~nm diode laser at about \SI{10}{mW} free-space power in the linear excitation regime. The PL emission from the SiV was collected using the same confocal setup, with a spatial resolution of around \SI{500}{nm}. The spectral distribution is analyzed with a Princeton Instrument HRS 750 spectrometer whose spectral resolution is around \SI{6}{GHz}, with integration times on the order of minutes. 
For dynamical strain measurements, the SiV emission is diluted in a broader optical bandwidth, so the corresponding integration time was increased further and thus susceptible to cosmic rays. To overcome this, multiple spectra with shorter time spans were taken, and cosmic rays were removed during data processing. 

\section{Mechanical characterization}

A homodyne interferometer setup, which was built into the confocal path described above (Fig.~\ref{Fig_SI_setup})~\cite{huang2025ultracoherent}, allows us to measure the mechanical motions of acoustic resonators by interfering light reflected from the vibrating resonator with a reference beam, converting motion-induced phase changes into measurable intensity signals. This enables characterization of the mechanical resonance frequencies (static strain), acoustic quality factors, and amplitude displacement (dynamical strain inferred) of our devices directly. In short, a TiSa laser at 730nm was split into a signal (sent to measure the device, about \SI{30}{\micro\watt}) and a local oscillator arm. After reflecting off the device, the signal and local oscillator arm are recombined at a balanced photodetector, where mechanical motion is read out in either the homodyne or heterodyne setting. 

Additionally, devices were driven through acceleration force with a piezoelectric actuator mounted to the back of the sample holder to excite and more easily identify mechanical modes. By sweeping the actuator drive frequency and monitoring the resulting displacement amplitude, we can identify the mechanical resonance frequency either through driven motion or its thermal motion. For the experiments where SiVs are strongly modulated, the displacement amplitude of the diamond string can approach \SI{500}{nm}, deep into the Duffing regime. We typically observe about a 2-3\% increase in resonance frequency due to stiffening. To mitigate the bistability and nonlinear phenomena, we approach the resonance from the lower-frequency side slowly using a waveform generator with the maximum driving voltage, right before the generation of acoustic combs shown in Fig.~\ref{FigSI_comb}. We observe that the driven motion is stable over 8 hours of operation. 

\section{Local static strain calibration using SiV PL spectra}
\begin{figure}[h]
    \centering
    \includegraphics[width=\linewidth]{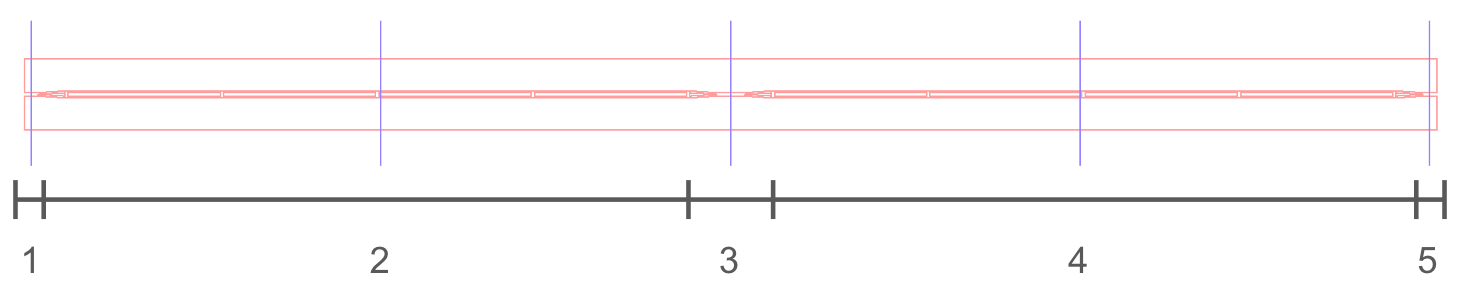}
    \caption{Device outline (red) with SiV implantation spots (purple). Due to the nature of the quasi-isotropic undercut, each segment with numbered labels could have a potential inhomogeneous cross-sectional area, leading to systematic errors in the estimated local static strain that the SiVs experience. We use the SiV PL spectra at each location across different average static stress to calibrate the relative static stress ratios among different sections, allowing us to accurately retrieve the real static stress at each SiV location even under non-uniform beam geometries. }
    \label{FigA2}
\end{figure}

Though the static strain experienced by the SiVs in the devices presented here is independent of implantation depth, which is an improvement over cantilever geometries, small thickness variations arising from the quasi-isotropic etching process could introduce local strain variations along the resonators. In this section, we discuss how we retrieve the local static strain at each SiV implantation spot to mitigate this systematic error. 

The strategy is the following: by comparing the PL spectra of SiVs at different locations as a function of average stress, retrieved from the resonator frequency, we can observe differences in how sensitive the SiV optical lines are to the average stress, due to local variations of cross-sections. This manifests in relative ratios among the raw-calibrated SiV susceptibilities using SiVs from different locations. Here we pick the SiVs at location 3 as our reference point, where the scaling factor $\beta_i$ of the local stress $\sigma_i$ at other locations $i$ can be used to calculate stress at each location by $\sigma_i = \beta_i \sigma_3$. Here, $\beta_i$ can be directly retrieved by the raw-calibrated SiV susceptibility ratios.

Next, we need to relate the local stress $\sigma_i$ to the global average stress $\sigma_\mathrm{eff}$. Here, we use the simplified model of a flexural wave propagating on a one-dimensional string. The velocity $v$ of a flexural wave on a tensioned string is given by $v = \sqrt{\sigma/\rho}$, where $\sigma$ is the local static stress and $\rho$ is the material density. The frequency $f$ of a resonant mode of a tensioned string is given by $f = (2L/v_\mathrm{eff})^{-1}$, where $L$ is the string length. This allows us to connect the averaged stress to the local stress using Rayleigh's quotient in a vibrating string,
\[\sigma_\mathrm{eff} \approx \frac{L}{2\int_0^L dx\sin^2(\pi x/L)/\sigma(x)}\]

To simplify this calculation, we segment the device into 5 different parts (Fig.~\ref{FigA2}). Due to the symmetry of the device, spots 1 and 5 are equivalent, and spots 2 and 4 are equivalent. Because sections 2 and 4 are the longest, while section 3 contributes more,  we obtain an approximated expression for the overall effective strain, 
\begin{equation*}
    {\sigma_\mathrm{eff}}^{-1} \approx 2\alpha_2{\sigma_2}^{-1}+2\alpha_3{\sigma_3}^{-1}
\end{equation*}
where $\sigma_\mathrm{eff}$ is the effective strain, $\alpha_i$ is a weight based on the length of each segment, and $\sigma_i$ is the local strain at each SiV location. Considering the relative stress ratios $\beta_i$ obtained from PL measurements, 
\begin{equation*}
    {\sigma_\mathrm{eff}}^{-1} \approx (2\alpha_2\beta_2^{-1}+2\alpha_3){\sigma_3}^{-1}
\end{equation*}
which allows us to first retrieve the local static stress at spot 3. Then the multiplicative factors $\beta_i$ allow us to obtain the local static stress at other locations. The strain was then calibrated based on this procedure for each spot group to obtain the colormap plotted in Fig.~\ref{Fig2}. By fitting the map with the literature susceptibility multiplied by a rescaling factor, we get a more accurate SiV strain susceptibility for its orbitals.  

\section{Benchmarking against orbital ground-state lifetime results}
In the main text, we obtained an SiV strain susceptibility rescaling factor of 0.356 from the SiV PL response to static strain. To validate this finding, we retrieve a rescaling factor of 0.370 using an independent method based on the temperature dependence of the orbital relaxation rate reported for bulk-diamond SIV defect centers in Ref.~\cite{kuruma2025controlling}. Instead of relying on the observed static shifts, the relaxation measurement reveals SiV sensitivity through resonant single-phonon emission at the frequency of ground-state orbital splitting. The agreement therefore provides more evidence on the correction factor, given that these two methods differ substantially in their systematics. In the following, we describe the analysis procedure leading to the correction factor. 

We treat the orbital ground state as a two-level system coupled to the bulk acoustic phonon bath, with single-phonon coupling rate $\Gamma_0$. The population re-equilibrium rate is $\Gamma(T)=\Gamma_0 \coth{(\Delta_\mathrm{GS}/2k_B T)}$, where $\omega_\mathrm{GS} = \Delta_\mathrm{GS}/\hbar$ is the ground state splitting. In the limit $k_BT\gg \Delta_\mathrm{GS}$, we obtain a linear temperature dependence $d\Gamma/dT=\Gamma_0(2k_B/\Delta_\mathrm{GS})$. We calibrate the susceptibility against this slope rather than a single lifetime, so that temperature-independent background channels are removed from the analysis, leaving only the contribution from the phonon. 

Next, we need to connect the spontaneous emission rate $\Gamma_0$ to the properties of SiV in diamond. This is modeled by a two-level system coupling to the bulk acoustic waves in an isotropic elastic continuum, by neglecting the surface modification of the phonon mode density. In the original degenerate doublet orbital states $|e_x\rangle$ and $|e_y\rangle$, the $E_g$ strain Hamiltonian that changes ground-state energy splitting is 
\[H_{\mathbf{\epsilon}} = h\begin{pmatrix}-E_x & E_y\\ E_y & E_x\end{pmatrix}, \]
where $E_x = d(\epsilon_{xx}-\epsilon_{yy}) + f\epsilon_{zx}$, $E_y = -2d\epsilon_{xy} + f\epsilon_{yz}$. Considering strain coupling to a unity amplitude strain wave with propagation direction $\hat n$ and polarization $\hat e_p$, the transition matrix element can be evaluated by $M_p(\hat n) = \langle g | H_{\hat \epsilon}|e\rangle$, where $\hat \epsilon_{ij}=(n_i e_{p,j} + n_j e_{p,i})/2$ is the unity strain tensor element. The single-phonon strain coupling rate can be analytically evaluated as
\[\Gamma_0 = \frac{\omega_\mathrm{GS}^3}{2\pi\rho\hbar}\left[\frac{\langle M_l(\hat n)\rangle_{\hat{n}}}{v_l^5}+\frac{\langle M_{t_1}(\hat n)\rangle_{\hat{n}}+\langle M_{t_2}(\hat n)\rangle_{\hat{n}}}{v_t^5}\right]\]
where $\rho=\SI{3515}{kg/m^3}$, $v_l=\SI{1.75e4}{m/s}$ and $v_t = \SI{1.28e4}{m/s}$. The averaging over direction is randomly sampled through all directions. From here, we can obtain the rescaling factor based on the high-temperature slope $\alpha = \SI{0.68}{MHz/K}$, where
\[s = \sqrt{2\pi\alpha \Delta_\mathrm{GS}/2k_B\Gamma_0(s=1)}=0.370.\]
Note that in this analysis $s$ is applied to $d\rightarrow sd$ and $f\rightarrow sf$. 

\section{Acoustic quality factor after ion implantation of Si}
\begin{figure}
    \centering\includegraphics[width=1\linewidth]{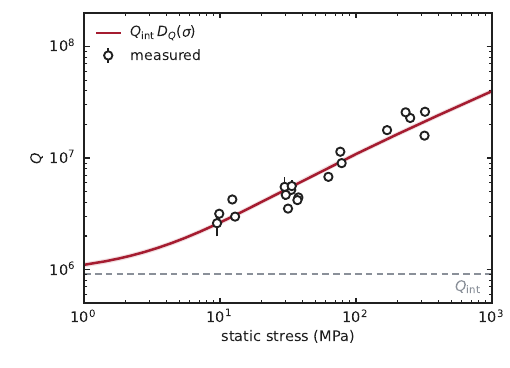}
    \caption{Measured mechanical quality factor versus static stress for the diamond strings (200~µm long,217~nm thick), with $Q = Q_\mathrm{int} D_Q(\sigma)$ using the simulated dilution factor $D_Q$ (1~std uncertainty). Fitted $Q_\mathrm{int} = (0.91 \pm 0.05) \times 10^6$, consistent with Ref~\cite{huang2025ultracoherent}, showing no implantation-induced dissipation at 4~K.}
    \label{FigSI_Qm}
\end{figure}

The $Q_\mathrm{int}$ before the implantation is summarized in Ref~\cite{huang2025ultracoherent}. The $Q_\mathrm{int}$ after Si implantation is extracted by fitting the dissipation dilution curve shown in Fig.~\ref{FigSI_Qm}, and is about $Q_\mathrm{int} = (0.91 \pm 0.05) \times 10^6$ at \SI{4}{K}. We find that this $Q_\mathrm{int}$ at 217~nm thickness and 200~nm width is consistent within 20\% deviation compared to Ref~\cite{huang2025ultracoherent}, where $Q_\mathrm{int}$ is \SI{1.3e6}{} at 100~nm thickness and 500~nm width, considering a 10\% difference in surface-volume ratios. Therefore, we believe experimentally there is no evidence that Si implantation leads to a noticeable increase in acoustic dissipation. 

Simulated SiV-induced relaxation damping due to dynamical strain coupling to its ground-state splitting is about $Q_\mathrm{int,GS}^{-1}\sim$\SI{e-9}{} per SiV at the clamp, and $Q_\mathrm{int,GS}^{-1}\sim$\SI{e-11}{} per SiV at the string center, both peaking around \SI{2}{K}. This means that theoretically we also don't expect this level of SiV concentration to induce noticeable acoustic damping across all temperature ranges.  
\section{SiV spectra under AC modulation}

\begin{figure*}
    \centering\includegraphics[width=1\linewidth]{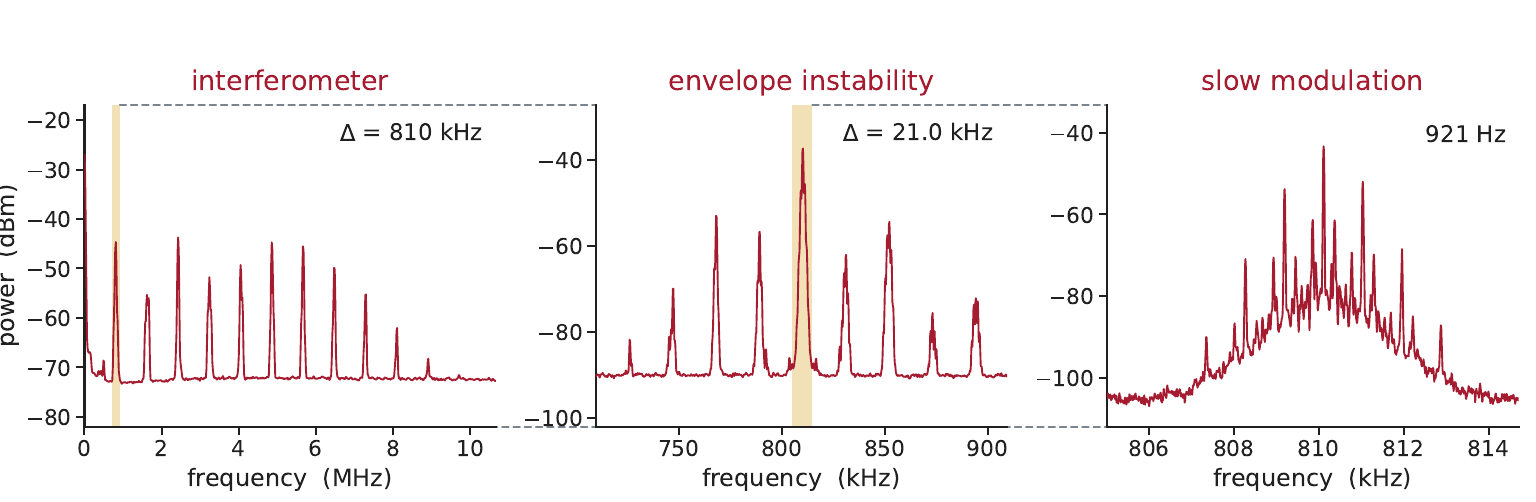}
    \caption{Presence of nested phononic frequency comb of a strongly driven diamond string, measured with the optical interferometer. The fundamental out-of-plane mode is externally excited, and Duffing hardened to \SI{810}{kHz}, which shows up in harmonics of the nonlinear transduction sidebands in the interferometer signal. Near the driving frequency, smaller sidebands spaced by \SI{21}{kHz} form a mechanical comb that indicates strong temporal modulation of the acoustic motion. We found that the comb formation is likely assisted by the nonlinear excitation of the third-order out-of-plane flexural mode, due to tension-mediated coupling between them. At higher resolution, we observe a slower modulation at \SI{921}{Hz}, possibly coming from the nonlinear potential associated with a driven Duffing oscillator.  }
    \label{FigSI_comb}
\end{figure*}

In the experiment that extracts $g_\mathrm{orbit}$, sometimes we need to externally excite the mechanical resonator mode to a large displacement amplitude, deep into the Duffing nonlinear regime. We observe that sometimes the SiV PL spectrum shows a shape inconsistent with a single-tone sine modulation. 

Upon carefully inspecting the interferometer signal of mechanical motion, we find that stochastically the mechanical motion goes into a nonlinear regime, which shows strong temporal modulation~\cite{Ochs22} of the driven mechanical motion as shown in Fig.~\ref{FigSI_comb}. As a driven Duffing resonator with linear damping can't undergo the Hopf bifurcation, we find that the frequency comb formation is assisted by the nonlinear excitation of the third-order out-of-plane flexural mode that approaches the 1:3 frequency ratio under Duffing hardening~\cite{Ganesan17,Qi20}. 

As we find that the stabilization of a nonlinear phononic state is challenging for minutes-long PL integration times in our device, we opted to study the SiV spectral response to two externally excited mechanical modes instead, where the temporal modulation should show similar characteristics compared to that of a nonlinear acoustic state. In the following, we analyze the difference in SiV spectral shapes under one-tone vs. two-tone modulation. 

Under a single-tone drive, the optical transition undergoes $\delta(t) = A \cos(\Omega_mt)$ modulation, with $\delta$ being the detuning from the unperturbed line and the $A$ the modulation amplitude proportional to the dynamical strain experienced by the SiV. As the mechanical frequency is far below the spectral resolution and the SiV optical response time, the accumulated PL spectrum measures the dwell-time density of the modulated SiV line, which is straightforwardly $p(\delta)\propto \sqrt{A^2-\delta^2}^{-1}$. Note that the line spends most of the time near its turning point as $\delta'(t)\rightarrow 0$, leading to the doublet splitting spaced by $2A$. When convoluting with a Lorentzian kernel of width $\gamma$ (spectrometer resolution), the density is of the form $S(\delta) \propto \mathrm{Im}\left[\sqrt{(\delta +i\gamma - A)(\delta +i\gamma + A)}^{-1/2}\right]$, which is used as the fitting function for our single-tone driven PL data. The kernel width is not used as a fitting parameter, and is fixed by a drive-off dataset. 

\begin{figure*}
    \centering\includegraphics[width=1\linewidth]{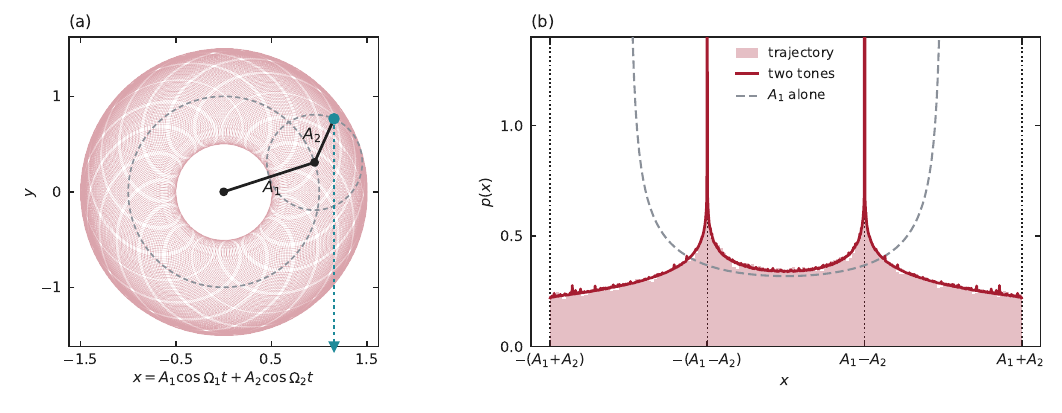}
    \caption{Dwell-time density of the SiV detuning under two-tone modulation. \textbf{(a)} The detuning is the x-projection of a point on an epicycle with a radius $A_1$ rotating at $\Omega_1$ that carries a radius $A_2$ rotating at $\Omega_2$. The trajectory is shown for $A_2/A_1 = 1/2$ and the ratio $\Omega_2/\Omega_1 = \pi$ over 100 cycles of $\Omega_1$. The trajectory fills the annulus between radii $A_1-A_2$ and $A_1+A_2$. \textbf{(b)} Density $p(x)$ of the projection compared with the single-tone density of $A_1$ alone. }
    \label{FigSI_twotone}
\end{figure*}

Under two drive tones, the detuning is a linear addition $\delta (t) = A_1 \cos(\Omega_1 t+ \phi_1) + A_2 \cos(\Omega_2 t+ \phi_2)$, where the measured frequency ratio $\Omega_2/\Omega_1=2.05$ for the first and second order flexural modes. Because both modes are actively driven, the modulation can be kept stable over hour-long data acquisition sessions. Because the relative phase between the two tones is linearly swept, the dwell time density is simply the convolution of the single-tone densities as $p(\delta = \delta_1 + \delta_2) = \int d\delta_1p_1(\delta_1)p_2(\delta - \delta_1)=\mathrm{FT}_{\xi\rightarrow \delta}\left[J_0(A_1\xi)J_0(A_2\xi)\right]$, which is evaluated numerically. 

Even though the end shape needs to be computed numerically, we can still understand the properties of this shape function intuitively, shown in Fig.~\ref{FigSI_twotone}. First, it's not a function of the drive frequency $\Omega_m$ as we assumed the phase relation between the two drives to be unstable, so the shape is only a function of $A_1$ and $A_2$. Because of the property of the convolution, the shape does not extend beyond $\delta=\pm(A_1+A_2)$, and diverges at $\delta = \pm(A_1-A_2)$ when the individual divergent peaks inside the convolution align with each other. The physics of this shape is the same mathematics as van Hove singularities in a 2D band structure~\cite{Van53}, where saddle points in phase space give rise to logarithmic divergence.

The fitting procedure minimizes the $\chi^2$ comparing four models: a free-width Lorentzian, one-tone shape, one-tone with a Gaussian blur, and two-tone shape. We found that the two-tone shape is always the minimum $\chi^2$ decisively by at least $\Delta\chi^2=86$, confirming that the SiV is indeed driven by two independent mechanical tones, whose dynamical strain ratio can be inferred by the SiV spectral shape. 

On the other hand, the optical interferometer allows us to estimate the dynamical strain ratio at the clamp independently, by placing the optical beam about 3-\SI{5}{\micro\meter} away from the clamping point. Even though the interferometer only measures displacement ratio $q_2/q_1$, near the clamping point $q_2/q_1 = G \epsilon_{\mathrm{d},2}/\epsilon_{\mathrm{d},1}$, where $G\approx 1.01$ and $\epsilon_{\mathrm{d}}$ is the dynamical strain. This will become evident in the next section, but intuitively this is because the mode shape is dominated by the boundary condition $u(x)\propto qk\ell(1-e^{-x/\ell})$, where the boundary layer length $\ell$ is the same between mechanical modes. 

Note that the mechanical signal at the boundary is really small, below that of the electrical crosstalk of the piezo drive. Here, we use the AOM to offset the local oscillator to the device reflection by a small frequency, enabling measurement of mechanical motion at a frequency different from that of the piezo drive. The errorbar includes the position uncertainty contribution around 3-\SI{5}{\micro\meter} location offset, as well as the variance of the ratio measured using different types of estimation, e.g., area sum vs. peak height. 

\section{Orbital-phonon coupling rates of dissipation-diluted beams}
\begin{figure*}
    \centering\includegraphics[width=1\linewidth]{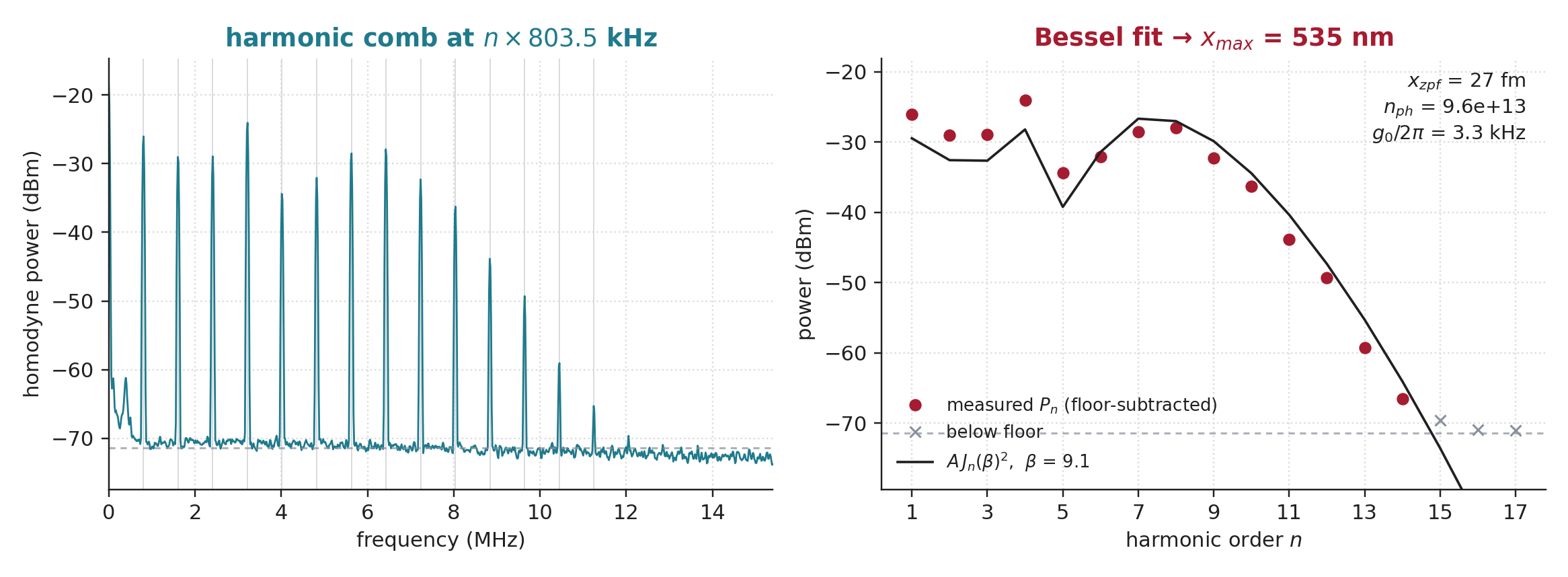}
    \caption{Interferometric displacement calibration. Left: homodyne spectrum of an excited acoustic mode at \SI{803.5}{kHz}, showing the harmonic comb at multiples of the drive frequency. Right: Harmonic sideband powers fitted with $P_n \propto J_n(\beta)^2$, determining the phase-modulation depth $\beta$ and the antinode displacement $x_\mathrm{max}=\beta\lambda/4\pi$, where $\lambda$ is the laser wavelength.}
    \label{FigSI_drive}
\end{figure*}
In the main manuscript, we demonstrate that static strain directly enhances dissipation dilution $D_Q>20$ of mechanical flexural modes by introducing the lossless tension energy along the string, and indirectly enhances orbit-mechanical coupling (ten-fold) of SiV defect centers near the resonator clamp by localizing the dynamical strain at the clamp. 

The dynamic strain localizes because the clamped boundary condition forces the tensioned string to bend within a boundary layer of length $\ell=h/\sqrt{12\epsilon}$, where $h$ is the thickness and $\epsilon$ is the static strain, such that there is no abrupt change of the gradient at the clamp boundary~\cite{fedorov2019generalized}. The standing-wave mode shape is therefore $u(x) = q \left[\sin(kx)-k\ell\left(1-e^{-x/\ell}\right)\right]$ with maximal displacement $q$, where the dynamical strain $\epsilon_\mathrm{d}(x) = -z_0u''(x)$, at $z_0\sim\SI{40}{nm}$ depth above the symmetry plane, has the following expression evaluated at the beam center and the beam clamp ($x_0$ offset distance),
\begin{gather*}
    |\partial \epsilon_\mathrm{d}/\partial q|_\mathrm{center} = z_0k^2\\
    |\partial \epsilon_\mathrm{d}/\partial q|_\mathrm{clamp} = z_0\left|k^2 \sin(kx_0) - \frac{k}{\ell}e^{-x_0/\ell}\right|
\end{gather*}
leading to a maximal dynamical strain enhancement factor around $D_\epsilon = \sqrt{12\epsilon}/kh$, which is evaluated to be around 15 for the best device in this study. 

The orbital coupling rate is evaluated as $g_\mathrm{orbit} = q_\mathrm{zpf}\partial \nu/\partial q$, where zero-point displacement $q_\mathrm{zpf} = \sqrt{\hbar/2m_\mathrm{eff}\Omega_m}$, mechanical frequency $\Omega_m = \sqrt{\frac{E}{\rho}\left(\epsilon k^2 + h^2k^4/12\right)}$, $k = \pi/L$. Experimentally, this number is straightforward to characterize. We can obtain the frequency modulation $\delta\nu$ of SiV optical lines by fitting the PL spectrum, whose resolution kernel is the Lorentzian shape of the non-driven SiV lines. We obtain the dynamical displacement $\delta q$ by fitting the interferometer spectrum consisting of a comb of harmonics from the homodyne transduction, whose sideband power $P_n \propto J_n(4\pi \delta q/\lambda)^2$ allows us to characterize $\delta q$ precisely using the laser wavelength $\lambda = \SI{740}{nm}$, see Fig.~\ref{FigSI_drive}. 

For simulating the optical response of SiV under dynamical displacement, we model the dynamical strain under the uniaxial assumption along the beam axis [110] with Poisson ratio $\nu_P = 0.1$. Rotating into the SiV orientation allows us to evaluate the coupling numerically via perturbation of the static strain splitting (M=GS, ES) by $\Delta_\mathrm{M} = \hbar\sqrt{\lambda_M^2 + 4S_M^2(\epsilon+\epsilon_\mathrm{d})^2}$, as well as the mean shift $s_0$ of the zero-phonon line using literature value~\cite{meesala2018strain} and the scale factor, leading to
\[\frac{\partial\nu_{C,D}}{\partial \epsilon_d} = s_0 - h_\mathrm{ES} \pm h_\mathrm{GS},\quad h_M = \frac{2S_M^2\epsilon}{\sqrt{\lambda_M^2 + 4S_M^2\epsilon^2}}.\]

Notice that the splitting contributions $h_M\rightarrow 0$ vanish at zero static strain, leaving the mean shift $s_0$ as the only contribution. Under large strain, $h_M\rightarrow S_M$, and saturates at $S_M$. In the main text figure showing the simulated results of $g_\mathrm{orbit}$ as a function of stress, we evaluated it using $g_\mathrm{orbit} = \left|\frac{\partial\nu_{C,D}}{\partial \epsilon_d}\right|\left|\frac{\partial\epsilon_d}{\partial q}\right| q_\mathrm{zpf}$, as an envelope over the axial C, axial D and transverse D branches. We can't reliably distinguish these lines in the experiment, while the transverse C branch is not in the wavelength band of analysis. In evaluating the static strain experienced by the emitter, we use the local strain corrected by the cross-section ratio at each location, whereas the mechanical properties are determined by the mean strain and the measured mechanical frequencies. The uncertainty of the implantation depth $\sim\SI{50}{nm}$ (\SI{76.5}{keV}) and $\sim\SI{65}{nm}$ (\SI{96.5}{keV}) each has about $\sim\SI{10}{nm}$ spread from SRIM simulations, which is not accounted for by the theory lines. The span of the theory lines account for the possible optical lines as well as the two implantation depths of SiVs in our device. 

We note~\footnote{We thank Slava Dobrovitski for bringing this to our attention.} that in Ref.~\cite{meesala2018strain} there are different SiV frames and different formula for connecting the level splittings to the local strain tensors in those frames. We found that these frame choices do not always work with the different expressions of the splittings written in that reference. We also found that some of the frames are improperly defined, and different expressions contradict each other. Improper pairing of frames and expressions could lead to drastic changes to the SiV strain response. As an example, the ground state strain susceptibility to the axial device strain could change by a factor of three. By numerically screening different frames, directly stated or inferred from the reference, with the different splitting expressions, we found that the following two combiations work: (i) SiV frame of $x=[11\bar{2}]$, $y=[\bar{1}10]$, $z=[111]$ paired with Eq.(2), and (ii) SiV frame of $x=[\bar{1}10]$, $y=[\bar{1}\bar{1}2]$, $z=[111]$ paired with Eq.(6,7). We used, in this work, SiV frame of $x=[\bar{1}\bar{1}2]$, $y=[1\bar{1}0]$, $z=[111]$ paired with formula Eq.(B4) with a flipped sign $f$ parameter from the reference, which could also numerically reproduce the Fig.2 experimental data of that reference. 

\section{Simulated spin-phonon coupling rates of dissipation-diluted beams}

\begin{figure}[t]
    \centering\includegraphics[width=1\linewidth]{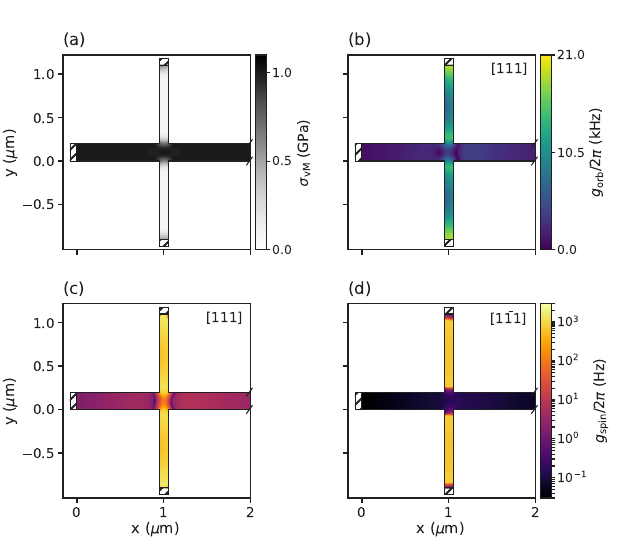}
    \caption{SiV spin-phonon coupling of a 200~µm diamond beam along [110] with a cross-shaped clamp (fundamental out-of-plane mode at 1.355~MHz, width 200~nm, thickness 100~nm). The beam is under 1~GPa tension, anchored through a 100~nm-wide transverse clamp placed 1~µm from the hard clamp (hatch pattern: fixed boundaries). \textbf{(a)} Static von Mises stress: the transverse clamp tethers remain strain-free while the main beam holds 1~GPa tension for dissipation dilution. \textbf{(b)} Dispersive orbital coupling per phonon for the [111] oriented SiV. \textbf{(c)}, \textbf{(d)} Spin–phonon coupling rates for the [111] and [1$\bar{1}$1] oriented SiV, with B = 0.355~T at ${62}^\circ$ to the SiV axis. The SiVs are located near the surface of the beam. }
    \label{FigSI_cross}
\end{figure}

Enhancement of spin-mechanical coupling is less straightforward, as spin-orbit hybridization is suppressed when there is an off-axis static strain field applied. Essentially, the same static strain that enhances the SiV ground-state splitting through symmetry breaking greatly suppresses the strain susceptibility of spin. 

However, one should recognize that static and dynamical strain components can be decoupled through engineering the clamp shape, so that SiV does not experience off-axis static strain (even no static strain at all), but experiences the same static-strain-localized off-axis dynamical strain that enhances spin-mechanical coupling. 

Here we demonstrate precisely one such design in simulation. We model a 200~µm diamond beam (width 200~nm, thickness 100~nm) along [110] with a cross-shaped clamp, where the beam is under 1~GPa tension with a fundamental out-of-plane mode at 1.355~MHz. The cross-shaped clamp is specifically to avoid static strain build-up in the transverse tethers, which would otherwise decouple spin from the ground state orbitals, while still enjoying the enhancement of dynamical strain localization near the clamps. 

We show in Fig.~\ref{FigSI_cross} the simulated spin-phonon coupling rate, where $B=\SI{0.355}{T}$ is applied at ${62}^\circ$ to the SiV axis, with spin frequency \SI{6.3}{GHz}. For the current device geometry, the optimized spin-phonon frequency-shift coupling rate $g_\mathrm{spin}/2\pi=\SI{1.5}{kHz}$ is obtained using the cross-clamp configuration, three orders of magnitude beyond that achievable by a simple doubly clamped string. This rate could be further improved by ten fold using a length of \SI{20}{\micro\meter} instead, following the scaling of $g_\mathrm{spin}\propto L^{-1}h^{-1/2}W^{-1/2}\sigma^{1/4}$, $C_\mathrm{sp}\propto L^{-1}  h^{-2}W^{-1}\sigma$. 

Using our current device geometry, assuming operation at 100~mK, spin coherence time $T_2 = 1/\gamma = \SI{1}{ms}$~\cite{Sukachev17} under dynamical decoupling at $\Omega_m/2\pi = \SI{1.355}{MHz}$, acoustic material quality factor $Q_\mathrm{int} = \SI{e8}{}$~\cite{huang2025ultracoherent}, we could obtain a maximum quantum cooperativity of $C_q=4g_\mathrm{spin}^2/\gamma \Gamma_\mathrm{th} = \SI{3e5}{}$, where acoustic thermal decoherence rate $\Gamma_\mathrm{th} = k_BT/\hbar D_QQ_\mathrm{int}=\SI{1.25}{s^{-1}}$. Here we use the bulk $T_1$ limit for the spin coherence with a misaligned B-field, due to enhanced strain coupling to the GHz phonon bath. This could be further improved by patterning GHz-phononic crystal structures around the SiV location to suppress the local density of modes. 

\section{Comparison to cantilever coupling}
A cantilever is the typical mechanical component for strain coupling to color centers, and therefore it is natural to compare it to the tensioned string we demonstrate here. The two geometries differ in where the bending occurs. Without tension, the fundamental modes of string and cantilever are similar, where the bending curvature decay constant is on the order of the beam length $L$. With the addition of tension, the bending is confined to the boundary layer for the tensioned beam, with the mode frequency also up-shifting due to the presence of the lossless tensile potential.

At equal length, the tensioned string's clamp strain per unit displacement exceeds the cantilever's by about $L/\ell$, which is about a 50-fold enhancement for our most strained device. However, the up-shifted frequency also leads to a lower zero-point displacement, such that the single-phonon strain is only enhanced by about six-fold. However, the cooperativity also benefits from mechanical-Q enhancement, which is about a factor of 30 in our demonstration. 

At equal frequency, the clamp strain per unit displacement is similar between the cantilever and the string. In this case, the cooperativity enhancement only comes from $D_Q$ and scales by $\sqrt{\epsilon}$ instead. 

\section{Comparison to other hybrid quantum system approaches}

One emerging architecture couples superconducting qubits, which exhibit fast gate operations but are limited by short coherence times, with long-lived acoustic modes hosted in mechanical resonators~\cite{yang2026mechanical, chu2017quantum}. 
Here, the superconducting qubits are commonly coupled to acoustic modes utilizing piezoelectric materials~\cite{chu2017quantum, han2020cavity}, which provide a conversion path between electrical and mechanical energy. However, intrinsic material defects and interfacial losses associated with heterogeneous material integration often degrade the quantum coherence of both systems~\cite{yang2023piezoelectric, gruenke2025surface}.

Heterogeneous integration of spin-mechanical coupling has been demonstrated in the presence of a strong magnetic field gradient~\cite{Kolkowitz12}. However, it has been found that non-contact friction~\cite{Arabmoheghi2026} leads to significant degradation of acoustic quality factors~\cite{Fung24} before an appreciable amount of coupling is reached. Therefore, similar approaches have evolved towards levitation utilizing rotational degrees of freedom~\cite{Delord2020}. 
Monolithic integration has thus far focused on strain-coupling and has realized efficient coupling of spins and orbitals to GHz~\cite{Joe2026} and MHz solid-state phonons~\cite{Ovartchaiyapong2014, Lee16}. 
Note that other than ion-implantation used in this work, approaches~\cite{Oh26} using delta-doping of NV centers also demonstrate the preservation of diamond's excellent acoustic properties, showing that monolithic integration does not evoke excess loss in diamond acoustics. 

% \clearpage
\bibliographystyle{apsrev4-2}
\bibliography{ref}

@article{kuruma2025controlling,
  title={Controlling interactions between high-frequency phonons and single quantum systems using phononic crystals},
  author={Kuruma, Kazuhiro and Pingault, Benjamin and Chia, Cleaven and Haas, Michael and Joe, Graham D and Assumpcao, Daniel Rimoli and Ding, Sophie Weiyi and Jin, Chang and Xin, CJ and Yeh, Matthew and others},
  journal={Nature Physics},
  volume={21},
  number={1},
  pages={77--82},
  year={2025},
  publisher={Nature Publishing Group UK London},
  doi={10.1038/s41567-024-02697-5},
  url={https://doi.org/10.1038/s41567-024-02697-5}
}

@article{huang2025ultracoherent,
  title={Ultracoherent self-assembled diamond nanomechanics reveals superfluid dynamics},
  author={Huang, Guanhao and Jin, Chang and Ding, Sophie Weiyi and Zhang, Chaoshen and Day, Aaron M and Elbs, Tobias and Sinclair, Neil and Joshi, Sukhad Dnyanesh and Defo, Rodrick Kuate and Halperin, Bertrand I and others},
  journal={arXiv preprint arXiv:2507.01217},
  year={2025},
  doi={10.48550/arXiv.2507.01217},
  url={https://arxiv.org/abs/2507.01217}
}

@article{kuruma2021coupling,
  title={Coupling of a single tin-vacancy center to a photonic crystal cavity in diamond},
  author={Kuruma, Kazuhiro and Pingault, Benjamin and Chia, Cleaven and Renaud, Dylan and Hoffmann, Patrick and Iwamoto, Satoshi and Ronning, Carsten and Lon{\v{c}}ar, Marko},
  journal={Applied Physics Letters},
  volume={118},
  number={23},
  year={2021},
  publisher={AIP Publishing},
  doi={10.1063/5.0051675},
  url={https://doi.org/10.1063/5.0051675}
}

@article{meesala2018strain,
  title={Strain engineering of the silicon-vacancy center in diamond},
  author={Meesala, Srujan and Sohn, Young-Ik and Pingault, Benjamin and Shao, Linbo and Atikian, Haig A and Holzgrafe, Jeffrey and G{\"u}ndo{\u{g}}an, Mustafa and Stavrakas, Camille and Sipahigil, Alp and Chia, Cleaven and others},
  journal={Physical Review B},
  volume={97},
  number={20},
  pages={205444},
  year={2018},
  publisher={APS},
  doi={10.1103/PhysRevB.97.205444},
  url={https://doi.org/10.1103/PhysRevB.97.205444}
}

@article{sohn2018controlling,
  title={Controlling the coherence of a diamond spin qubit through its strain environment},
  author={Sohn, Young-Ik and Meesala, Srujan and Pingault, Benjamin and Atikian, Haig A and Holzgrafe, Jeffrey and G{\"u}ndo{\u{g}}an, Mustafa and Stavrakas, Camille and Stanley, Megan J and Sipahigil, Alp and Choi, Joonhee and others},
  journal={Nature communications},
  volume={9},
  number={1},
  pages={2012},
  year={2018},
  publisher={Nature Publishing Group UK London},
  doi={10.1038/s41467-018-04340-3},
  url={https://doi.org/10.1038/s41467-018-04340-3}
}

@article{assumpcao2023deterministic,
  title={Deterministic creation of strained color centers in nanostructures via high-stress thin films},
  author={Assumpcao, Daniel R and Jin, Chang and Sutula, Madison and Ding, Sophie W and Pham, Phong and Knaut, Can M and Bhaskar, Mihir K and Panday, Abishrant and Day, Aaron M and Renaud, Dylan and others},
  journal={Applied Physics Letters},
  volume={123},
  number={24},
  year={2023},
  publisher={AIP Publishing},
  doi={10.1063/5.0171558},
  url={https://doi.org/10.1063/5.0171558}
}

@article{machielse2019quantum,
  title={Quantum interference of electromechanically stabilized emitters in nanophotonic devices},
  author={Machielse, Bartholomeus and Bogdanovic, Stefan and Meesala, Srujan and Gauthier, Scarlett and Burek, Michael J and Joe, Graham and Chalupnik, Michelle and Sohn, Young-Ik and Holzgrafe, Jeffrey and Evans, Ruffin E and others},
  journal={Physical Review X},
  volume={9},
  number={3},
  pages={031022},
  year={2019},
  publisher={APS},
  doi={10.1103/PhysRevX.9.031022},
  url={https://doi.org/10.1103/PhysRevX.9.031022}
}

@article{yang2026mechanical,
  title={Mechanical resonator--based quantum computing},
  author={Yang, Yu and Kladari{\'c}, Igor and Skrabulis, Martynas and Eichenberger, Michael and Marti, Stefano and Storz, Simon and Esche, Jonathan and Bell{\'e}s, Raquel Garc{\'\i}a and Kern, Max-Emanuel and Omahen, Andraz and others},
  journal={Science},
  volume={392},
  number={6801},
  pages={972--976},
  year={2026},
  publisher={American Association for the Advancement of Science},
  doi={10.1126/science.aef4139},
  url={https://doi.org/10.1126/science.aef4139}
}

@article{
doi:10.1126/science.adr2464,
author = {Yu Yang  and Igor Kladarić  and Maxwell Drimmer  and Uwe von Lüpke  and Daan Lenterman  and Joost Bus  and Stefano Marti  and Matteo Fadel  and Yiwen Chu },
title = {A mechanical qubit},
journal = {Science},
volume = {386},
number = {6723},
pages = {783-788},
year = {2024},
doi = {10.1126/science.adr2464},
URL = {https://www.science.org/doi/abs/10.1126/science.adr2464}}

@article{
doi:10.1126/science.aeh7535,
author = {Takuma Makihara  and Erik Szakiel  and Matthew P. Maksymowych  and Oliver A. Hitchcock  and Kaveh Pezeshki  and Rachel G. Gruenke-Freudenstein  and Mihir Pendharkar  and Shannon P. Harvey  and David I. Schuster  and Amir H. Safavi-Naeini },
title = {Quantum jumps of sound},
journal = {Science},
volume = {393},
number = {6817},
pages = {1217-1220},
year = {2026},
doi = {10.1126/science.aeh7535},
URL = {https://www.science.org/doi/abs/10.1126/science.aeh7535}}

@article{Qi20,
    author = {Qi, Zhen and Menyuk, Curtis R. and Gorman, Jason J. and Ganesan, Adarsh},
    title = {Existence conditions for phononic frequency combs},
    journal = {Applied Physics Letters},
    volume = {117},
    number = {18},
    pages = {183503},
    year = {2020},
    month = {11},
    issn = {0003-6951},
    doi = {10.1063/5.0025314},
    url={https://doi.org/10.1063/5.0025314}
}

@article{Ochs22,
  title = {Frequency Comb from a Single Driven Nonlinear Nanomechanical Mode},
  author = {Ochs, J. S. and Bone{\ss{}}, D. K. J. and Rastelli, G. and Seitner, M. and Belzig, W. and Dykman, M. I. and Weig, E. M.},
  journal = {Phys. Rev. X},
  volume = {12},
  issue = {4},
  pages = {041019},
  numpages = {15},
  year = {2022},
  month = {Nov},
  publisher = {American Physical Society},
  doi = {10.1103/PhysRevX.12.041019},
  url = {https://link.aps.org/doi/10.1103/PhysRevX.12.041019}
}

@article{Ganesan17,
  title = {Phononic Frequency Comb via Intrinsic Three-Wave Mixing},
  author = {Ganesan, Adarsh and Do, Cuong and Seshia, Ashwin},
  journal = {Phys. Rev. Lett.},
  volume = {118},
  issue = {3},
  pages = {033903},
  numpages = {5},
  year = {2017},
  month = {Jan},
  publisher = {American Physical Society},
  doi = {10.1103/PhysRevLett.118.033903},
  url = {https://link.aps.org/doi/10.1103/PhysRevLett.118.033903}
}

@article{Sukachev17,
  title = {Silicon-Vacancy Spin Qubit in Diamond: A Quantum Memory Exceeding 10 ms with Single-Shot State Readout},
  author = {Sukachev, D. D. and Sipahigil, A. and Nguyen, C. T. and Bhaskar, M. K. and Evans, R. E. and Jelezko, F. and Lukin, M. D.},
  journal = {Phys. Rev. Lett.},
  volume = {119},
  issue = {22},
  pages = {223602},
  numpages = {6},
  year = {2017},
  month = {Nov},
  publisher = {American Physical Society},
  doi = {10.1103/PhysRevLett.119.223602},
  url = {https://link.aps.org/doi/10.1103/PhysRevLett.119.223602}
}

@article{engelsen2024ultrahigh,
  title={Ultrahigh-quality-factor micro-and nanomechanical resonators using dissipation dilution},
  author={Engelsen, Nils Johan and Beccari, Alberto and Kippenberg, Tobias Jan},
  journal={Nature Nanotechnology},
  volume={19},
  number={6},
  pages={725--737},
  year={2024},
  publisher={Nature Publishing Group UK London},
  doi={10.1038/s41565-023-01597-8},
  url={https://doi.org/10.1038/s41565-023-01597-8}
}

@article{fedorov2019generalized,
  title={Generalized dissipation dilution in strained mechanical resonators},
  author={Fedorov, Sergey A and Engelsen, Nils J and Ghadimi, Amir H and Bereyhi, Mohammad J and Schilling, Ryan and Wilson, Dalziel J and Kippenberg, Tobias J},
  journal={Physical Review B},
  volume={99},
  number={5},
  pages={054107},
  year={2019},
  publisher={APS},
  doi={10.1103/PhysRevB.99.054107},
  url={https://doi.org/10.1103/PhysRevB.99.054107}
}

@article{kurizki2015quantum,
  title={Quantum technologies with hybrid systems},
  author={Kurizki, Gershon and Bertet, Patrice and Kubo, Yuimaru and M{\o}lmer, Klaus and Petrosyan, David and Rabl, Peter and Schmiedmayer, J{\"o}rg},
  journal={Proceedings of the National Academy of Sciences},
  volume={112},
  number={13},
  pages={3866--3873},
  year={2015},
  publisher={National Academy of Sciences},
  doi={10.1073/pnas.1419326112},
  url={https://doi.org/10.1073/pnas.1419326112}
}

@article{chu2017quantum,
  title={Quantum acoustics with superconducting qubits},
  author={Chu, Yiwen and Kharel, Prashanta and Renninger, William H and Burkhart, Luke D and Frunzio, Luigi and Rakich, Peter T and Schoelkopf, Robert J},
  journal={Science},
  volume={358},
  number={6360},
  pages={199--202},
  year={2017},
  publisher={American Association for the Advancement of Science},
  doi={10.1126/science.aao1511},
  url={https://doi.org/10.1126/science.aao1511}
}

@article{han2020cavity,
  title={Cavity piezo-mechanics for superconducting-nanophotonic quantum interface},
  author={Han, Xu and Fu, Wei and Zhong, Changchun and Zou, Chang-Ling and Xu, Yuntao and Sayem, Ayed Al and Xu, Mingrui and Wang, Sihao and Cheng, Risheng and Jiang, Liang and others},
  journal={Nature communications},
  volume={11},
  number={1},
  pages={3237},
  year={2020},
  publisher={Nature Publishing Group UK London},
  doi={10.1038/s41467-020-17053-3},
  url={https://doi.org/10.1038/s41467-020-17053-3}
}

@article{yang2023piezoelectric,
  title={Piezoelectric loss of superconducting microwave resonators integrated with thin-film lithium niobate},
  author={Yang, Likai and Xu, Yuntao and Li, Chunzhen and Xie, Jiacheng and Shen, Mohan and Tang, Hong X},
  journal={Physical Review Applied},
  volume={20},
  number={5},
  pages={054026},
  year={2023},
  publisher={APS},
  doi={10.1103/PhysRevApplied.20.054026},
  url={https://doi.org/10.1103/PhysRevApplied.20.054026}
}

@article{gruenke2025surface,
  title={Surface and bulk two-level-system losses in lithium niobate acoustic resonators},
  author={Gruenke-Freudenstein, Rachel G and Szakiel, Erik and Multani, Gitanjali P and Makihara, Takuma and Hayden, Akasha G and Khalatpour, Ali and Wollack, E Alex and Akoto-Yeboah, Antonia and Salmani-Rezaie, Salva and Safavi-Naeini, Amir H},
  journal={Physical Review Applied},
  volume={23},
  number={6},
  pages={064055},
  year={2025},
  publisher={APS},
  doi={10.1103/tgs3-kkw8},
  url={https://doi.org/10.1103/tgs3-kkw8}
}

@article{xiang2013hybrid,
  title={Hybrid quantum circuits: Superconducting circuits interacting with other quantum systems},
  author={Xiang, Ze-Liang and Ashhab, Sahel and You, JQ and Nori, Franco},
  journal={Reviews of Modern Physics},
  volume={85},
  number={2},
  pages={623--653},
  year={2013},
  publisher={APS},
  doi={10.1103/RevModPhys.85.623},
  url={https://doi.org/10.1103/RevModPhys.85.623}
}

@article{Oh26,
author = {Hyunseok Oh and Viraj Dharod and Carl Padgett and Lillian B. Hughes Wyatt and Jayameenakshi Venkatraman and Shreyas Parthasarathy and Ekaterina Osipova and Ian Hedgepeth and Jeffrey V. Cady and Luca Basso and Yongqiang Wang and Michael Titze and Edward S. Bielejec and Andrew M. Mounce and Dirk Bouwmeester and Ania C. Bleszynski Jayich},
journal = {Optica},
number = {3},
pages = {485--490},
publisher = {Optica Publishing Group},
title = {Spin-embedded diamond optomechanical resonator with a mechanical quality factor exceeding one million},
volume = {13},
month = {Mar},
year = {2026},
doi = {10.1364/OPTICA.577541},
url={https://doi.org/10.1364/OPTICA.577541}
}

@Article{Flühmann2019,
author={Fl{\"u}hmann, C.
and Nguyen, T. L.
and Marinelli, M.
and Negnevitsky, V.
and Mehta, K.
and Home, J. P.},
title={Encoding a qubit in a trapped-ion mechanical oscillator},
journal={Nature},
year={2019},
month={Feb},
day={01},
volume={566},
number={7745},
pages={513-517},
issn={1476-4687},
doi={10.1038/s41586-019-0960-6},
url={https://doi.org/10.1038/s41586-019-0960-6}
}

@article{Kolkowitz12,
author = {Shimon Kolkowitz  and Ania C. Bleszynski Jayich  and Quirin P. Unterreithmeier  and Steven D. Bennett  and Peter Rabl  and J. G. E. Harris  and Mikhail D. Lukin },
title = {Coherent Sensing of a Mechanical Resonator with a Single-Spin Qubit},
journal = {Science},
volume = {335},
number = {6076},
pages = {1603-1606},
year = {2012},
doi = {10.1126/science.1216821},
url={https://doi.org/10.1126/science.1216821}
}

@article{Fung24,
  title = {Toward Programmable Quantum Processors Based on Spin Qubits with Mechanically Mediated Interactions and Transport},
  author = {Fung, F. and Rosenfeld, E. and Schaefer, J. D. and Kabcenell, A. and Gieseler, J. and Zhou, T. X. and Madhavan, T. and Aslam, N. and Yacoby, A. and Lukin, M. D.},
  journal = {Phys. Rev. Lett.},
  volume = {132},
  issue = {26},
  pages = {263602},
  numpages = {8},
  year = {2024},
  month = {Jun},
  publisher = {American Physical Society},
  doi = {10.1103/PhysRevLett.132.263602},
  url = {https://link.aps.org/doi/10.1103/PhysRevLett.132.263602}
}

@Article{Arabmoheghi2026,
author={Arabmoheghi, Amirali
and Zicoschi, Alessio
and Arregui, Guillermo
and Bereyhi, Mohammad J.
and Xia, Yi
and Engelsen, Nils J.
and Kippenberg, Tobias J.},
title={Non-contact friction in ultracoherent nanomechanical resonators near dielectric materials},
journal={Nature Physics},
year={2026},
month={Sep},
day={01},
volume={22},
number={9},
pages={1506-1512},
issn={1745-2481},
doi={10.1038/s41567-026-03350-z},
url={https://doi.org/10.1038/s41567-026-03350-z}
}

@Article{Joe2026,
author={Joe, Graham
and Haas, Michael
and Kuruma, Kazuhiro
and Jin, Chang
and Kang, Dongyeon Daniel
and Ding, Sophie W.
and Chia, Cleaven
and Warner, Hana
and Pingault, Benjamin
and Machielse, Bartholomeus
and Meesala, Srujan
and Lon{\v{c}}ar, Marko},
title={Purcell-enhanced spin--phonon coupling with a single colour centre},
journal={Nature},
year={2026},
month={May},
day={01},
volume={653},
number={8114},
pages={378-383},
issn={1476-4687},
doi={10.1038/s41586-026-10495-7},
url={https://doi.org/10.1038/s41586-026-10495-7}
}

@Article{Ovartchaiyapong2014,
author={Ovartchaiyapong, Preeti
and Lee, Kenneth W.
and Myers, Bryan A.
and Jayich, Ania C. Bleszynski},
title={Dynamic strain-mediated coupling of a single diamond spin to a mechanical resonator},
journal={Nature Communications},
year={2014},
month={Jul},
day={18},
volume={5},
number={1},
pages={4429},
issn={2041-1723},
doi={10.1038/ncomms5429},
url={https://doi.org/10.1038/ncomms5429}
}

@article{Lee16,
  title = {Strain Coupling of a Mechanical Resonator to a Single Quantum Emitter in Diamond},
  author = {Lee, Kenneth W. and Lee, Donghun and Ovartchaiyapong, Preeti and Minguzzi, Joaquin and Maze, Jero R. and Bleszynski Jayich, Ania C.},
  journal = {Phys. Rev. Appl.},
  volume = {6},
  issue = {3},
  pages = {034005},
  numpages = {17},
  year = {2016},
  month = {Sep},
  publisher = {American Physical Society},
  doi = {10.1103/PhysRevApplied.6.034005},
  url = {https://link.aps.org/doi/10.1103/PhysRevApplied.6.034005}
}

@Article{Delord2020,
author={Delord, T.
and Huillery, P.
and Nicolas, L.
and H{\'e}tet, G.},
title={Spin-cooling of the motion of a trapped diamond},
journal={Nature},
year={2020},
month={Apr},
day={01},
volume={580},
number={7801},
pages={56-59},
issn={1476-4687},
doi={10.1038/s41586-020-2133-z},
url={https://doi.org/10.1038/s41586-020-2133-z}
}

@Article{Brock2025,
author={Brock, Benjamin L.
and Singh, Shraddha
and Eickbusch, Alec
and Sivak, Volodymyr V.
and Ding, Andy Z.
and Frunzio, Luigi
and Girvin, Steven M.
and Devoret, Michel H.},
title={Quantum error correction of qudits beyond break-even},
journal={Nature},
year={2025},
month={May},
day={01},
volume={641},
number={8063},
pages={612-618},
issn={1476-4687},
doi={10.1038/s41586-025-08899-y},
url={https://doi.org/10.1038/s41586-025-08899-y}
}

@article{Penrose2014,
	author = {Roger Penrose},
	doi = {10.1007/s10701-013-9770-0},
	journal = {Foundations of Physics},
	number = {5},
	pages = {557--575},
	publisher = {Springer Us},
	title = {On the Gravitization of Quantum Mechanics 1: Quantum State Reduction},
	volume = {44},
	year = {2014},
	url={https://doi.org/10.1007/s10701-013-9770-0}
}

@article{Van53,
  title = {The Occurrence of Singularities in the Elastic Frequency Distribution of a Crystal},
  author = {Van Hove, L{\'e}on},
  journal = {Phys. Rev.},
  volume = {89},
  issue = {6},
  pages = {1189--1193},
  numpages = {0},
  year = {1953},
  month = {Mar},
  publisher = {American Physical Society},
  doi = {10.1103/PhysRev.89.1189},
  url = {https://link.aps.org/doi/10.1103/PhysRev.89.1189}
}

@Article{Youssefi2023,
author={Youssefi, Amir
and Kono, Shingo
and Chegnizadeh, Mahdi
and Kippenberg, Tobias J.},
title={A squeezed mechanical oscillator with millisecond quantum decoherence},
journal={Nature Physics},
year={2023},
month={Nov},
day={01},
volume={19},
number={11},
pages={1697-1702},
issn={1745-2481},
doi={10.1038/s41567-023-02135-y},
url={https://doi.org/10.1038/s41567-023-02135-y}
}

@Article{Rabl2010,
author={Rabl, P.
and Kolkowitz, S. J.
and Koppens, F. H. L.
and Harris, J. G. E.
and Zoller, P.
and Lukin, M. D.},
title={A quantum spin transducer based on nanoelectromechanical resonator arrays},
journal={Nature Physics},
year={2010},
month={Aug},
day={01},
volume={6},
number={8},
pages={602-608},
issn={1745-2481},
doi={10.1038/nphys1679},
url={https://doi.org/10.1038/nphys1679}
}

\end{document}